\documentclass[review,11pt]{elsarticle}
\biboptions{sort&compress}
\usepackage{iftex}
\ifluatex
\pdfvariable minorversion 6
\fi
\ifpdftex
\fi

\usepackage{placeins}
\usepackage{epsfig}
\usepackage{bm}
\usepackage{hyperref}
\usepackage{amsmath,amssymb,amsfonts,amsthm,bm}
\usepackage{cancel}
\usepackage{float}
\usepackage{miller}
\usepackage{subcaption}
\usepackage{booktabs}
\usepackage[margin=1in]{geometry}
\usepackage{algpseudocode}
\usepackage{soul}
\usepackage{todonotes}
\usepackage{color}

\definecolor{ic}{HTML}{FF4047}
\definecolor{ea}{HTML}{FF6347}

\graphicspath{{./}{./fig_drafts/}}

\begin{document}

\begin{frontmatter}

\title{The structure and migration of twin boundaries in tetragonal \texorpdfstring{$\beta$-Sn}{beta-Sn}: an application of machine learning based interatomic potentials}

\author[XCP5]{Ian Chesser\texorpdfstring{\corref{mycorrespondingauthor}}{}}
\author[MPA]{Mashroor Nitol}
\author[UCI]{Esther C. Hessong}
\author[UIUC]{Himanshu Joshi} 
\author[UIUC]{Nikhil Admal}
\author[IA]{Brandon Runnels}
\author[XCP5]{Daniel N. Blaschke}
\author[MST]{Khanh Dang}
\author[XCP5]{Abigail Hunter}
\author[MPA]{Saryu Fensin}

\cortext[mycorrespondingauthor]{Corresponding author}

\address[XCP5]{XCP-5, Los Alamos National Laboratory, Los Alamos, New Mexico, NM USA}
\address[MPA]{MPA-CINT, Los Alamos National Laboratory, Los Alamos, New Mexico, NM USA}
\address[UCI]{Department of Materials Science and Engineering, University of California, Irvine, CA USA}
\address[UIUC]{Department of Mechanical Science and Engineering, University of Illinois at Urbana-Champaign, Urbana, IL USA}
\address[IA]{Department of Aerospace Engineering, Iowa State University, Ames, IA USA}
\address[MST]{MST-8, Los Alamos National Laboratory, Los Alamos, New Mexico, NM USA}

\date{}

\begin{abstract}
Although atomistic simulations have contributed significantly to our understanding of twin boundary structure and migration in metals and alloys with hexagonal close packed (HCP) crystal structures, few direct atomistic studies of twinning have been conducted for other types of low symmetry materials, in large part due to a lack of reliable interatomic potentials. In this work, we examine twin boundary structure and migration in a tetragonal material, $\beta$-Sn, comparing high resolution Transmission Electron Microscopy (TEM) images of deformation twins in $\beta$-Sn to the results of direct atomistic simulations using multiple interatomic potentials. ML-based potentials developed in this work are found to give results consistent with our experimental data, revealing faceted twin boundary structures formed by the nucleation and motion of twinning disconnections. We use bicrystallographic methods in combination with atomistic simulations to analyze the structure, energy and shear coupled migration of observed twin facets in $\beta$-Sn. In analogy to Prismatic-Basal (PB/BP) interfaces in HCP metals, we discover low energy asymmetric Prismatic-A-plane (PA/AP) interfaces important to twin growth in $\beta$-Sn. A Moment Tensor Potential (MTP) and Rapid Artificial Neural Network (RANN) interatomic potential suitable for studying twinning and phase transformations in Sn are made publicly available as part of this work.  

\end{abstract}

%

\end{frontmatter}


\pagebreak
\tableofcontents

\section{Introduction}
\label{sec:intro}

Twinning in metallic tin (Sn) is significant from both a practical and theoretical standpoint. $\beta$-Sn, the thermodynamically stable phase under ambient conditions, has a low symmetry body-centered tetragonal (BCT) crystal structure that cannot easily accommodate slip under arbitrary loading conditions. In a 1932 Nature paper \cite{chalmers1932cry}, Chalmers et al. remarked that ``it is an observation of antiquity that when a bar of tin is bent, it emits a characteristic creaking, known as the cry of tin''. The rapid nucleation and growth of deformation twins is thought to be the origin of tin's cry \cite{chalmers1932cry, tu1970direct, christian2002}, though the atomic scale details of this process remain obscure. 

Twinning is an especially important deformation mechanism in  $\beta$-Sn during low temperature or high strain rate deformation \cite{tu1970direct, yang2007deformation, minor2014, pokharel2023, nguyen2024calibration, PTW2025}. Below the ductile-to-brittle transition temperature of 150 K, twinning in $\beta$-Sn is found to be the dominant deformation mechanism \cite{minor2014}. Twinning-induced embrittlement is of concern in the electronics industry where Sn-alloy based solder joints may fail via cracking along twin boundaries under thermal cycling \cite{minor2014, kaira2016microscale}. Apart from its applications in electronics, Sn has become a standard material for studying multiphase thermomechanical behavior due its low melting point (505 K) and accessible high pressure phase transformations \cite{nguyen2024calibration}. Profuse twinning is observed during impact loading of $\beta$-Sn \cite{yang2007deformation, lim2022characterization} and there is a need to integrate twinning into multiphase strength models for Sn \cite{nguyen2024calibration, PTW2025}. 

At the atomic scale, needle-like deformation twins in $\beta$-Sn are delineated by low-energy, special grain boundaries (GBs) referred to as coherent twin boundaries (CTBs) along their long axes. These CTBs not only appear during deformation twinning in $\beta$-Sn, but are also the most commonly observed GBs in cast or electrodeposited $\beta$-Sn microstructures \cite{rowenhorst2005measurements, telang2005orientation, kaira2016microscale,  reeve2018beta, gupta2022electrodeposition}. CTBs in $\beta$-Sn are thought to have the lowest excess energies among all GBs in $\beta$-Sn \cite{rowenhorst2005measurements, ware2022}, explaining their common occurrence; however, CTB energies in $\beta$-Sn have not been precisely measured. Twin-related networks of GBs in $\beta$-Sn provide barriers to rapid atomic diffusion, suppressing deleterious diffusion-related phenomena such as whisker growth \cite{Tu:2005, Buchovecky:2009, Jagtap:2021}. The two twin systems most commonly observed in $\beta$-Sn  are the \hkl(301)\hkl[-103] and \hkl(101)\hkl[-101] conjugate twins \cite{tu1970direct, christian2002}. \hkl(301)-type twins are reported to be more common than  \hkl(101)-type twins across deformation and solidification experiments \cite{tu1970direct, yang2007deformation, gupta2022electrodeposition}, but no explanation for a bias in twin variant selection has been given. A detailed description of twin boundary crystallography in $\beta$-Sn is given in Section \ref{sec:twin_growth}. 

Despite the common experimental observation of twinning and CTBs in $\beta$-Sn, the relaxed structures and growth mechanisms of twin nuclei in $\beta$-Sn remain unknown. On the other hand, since the 1980s, atomistic simulations with atomic resolution have been transformative to our understanding of deformation and annealing twins in metals with cubic and HCP crystal structures \cite{beyerlein2014growth, hirth2016disconnections, yue2023twin}. As one example, the 3D structure of the most commonly observed type of deformation twin in HCP Mg has been characterized across molecular dynamics (MD) simulations and experiments \cite{liu2019three}, leading to more accurate models for twin growth \cite{gong2021effects, runnels2025} and the discovery of unexpected twin boundary facets and migration mechanisms \cite{dang2020formation, yue2023twin}. Atomistic simulations enable high throughput discovery-scale studies of twinning which help guide more costly high resolution experiments and inform mesoscale models of twin growth.  

A critical ingredient of atomistic simulations of twinning is an accurate interatomic potential. For metals and metallic alloys, the classical Embedded-Atom Method (EAM) \cite{EAM, EAM2} and its extension, the Modified Embedded-Atom Method (MEAM)\cite{MEAM}, are widely used in studies of twinning in metals like Cu or Mg \cite{mishinCu, liuMg, wang2010detwinning, liu2019three, wang2012atomic}. These potentials approximate atomic interactions via a physically inspired functional form that includes many-body effects, making them well-suited for modeling metallic bonding. Ko \textit{et al.} fit a MEAM potential that captures elastic constants and lattice parameters within the $\beta$-Sn phase reasonably well compared to Density Functional Theory (DFT) data \cite{Ko:2018}. 

While EAM and MEAM potentials have been successful in modeling twinning in many materials, they struggle to accurately describe metals with more complex bonding characteristics, such as $\beta$-Sn, which exhibits mixed covalent-metallic bonding. Over the past decade, a new class of machine learning (ML)-based potentials have emerged, offering a more flexible and accurate framework for capturing atomic interactions in low symmetry materials with mixed bonding character \cite{zuo2020performance, mishin2021machine}. An ML-based Sn potential was recently developed to study the multiphase strength of Sn, capturing the phase diagram of Sn and defect-related properties of each phase better than any prior potential \cite{Nitol:2023}. Interatomic potentials for $\beta$-Sn have not yet been tested for their ability to capture twinning. 

In this work, we study the structure and migration of twin boundaries in $\beta$-Sn via a combination of experiments, atomistic simulations and bicrystallography-based methods. The ML-based potentials developed in this work are found to accurately capture experimentally observed twinning shears and structures for both \hkl(301)-type and \hkl(101)-type twins. Three scientific contributions of this work are the measurement of energies for common twin-related interfaces in $\beta$-Sn, the characterization of disconnection-mediated twin boundary migration mechanisms and the discovery of asymmetric facets important to twin growth in $\beta$-Sn.

\section{Methods}
\label{sec:methods}

\noindent The Large-scale Atomic/Molecular Massively Parallel Simulator
(LAMMPS) \cite{LAMMPS} was utilized to conduct molecular statics (MS) and MD simulations. All DFT calculations were performed using 
VASP\cite{hafner2008ab} (version 6.3.2) within the generalized gradient approximation (GGA)\cite{perdew1996generalized} 
Perdew-Burke-Ernzerhof (PBE) framework. A plane-wave cutoff 
energy of 520~eV and a $K$-point spacing of $2\pi \times 0.01$~\AA$^{-1}$ were 
employed across all structural configurations in this work. 

The software package OVITO
\cite{OVITO} was used to visualize atomic structures and dynamics. In particular, the Polyhedral Template Matching (PTM) modifier \cite{PTM} was used to distinguish between bulk and GB atoms in $\beta$-Sn structures with a root mean-squared cutoff of 0.25 (unless otherwise noted). Bulk atoms have approximately simple cubic coordination and are colored purple while all other colors represent disordered local environments. The approximately simple cubic coordination of individual atoms in bulk $\beta$-Sn is a consequence of its atomic basis within the BCT lattice. $\beta$-Sn has a multilattice structure with two interpenetrating BCT lattices separated by a translation of \hkl[a/2, 0, c/4]. 

\subsection{Experimental Methods}

\begin{figure}[H]
\centering\leavevmode \includegraphics[width=0.95\textwidth]{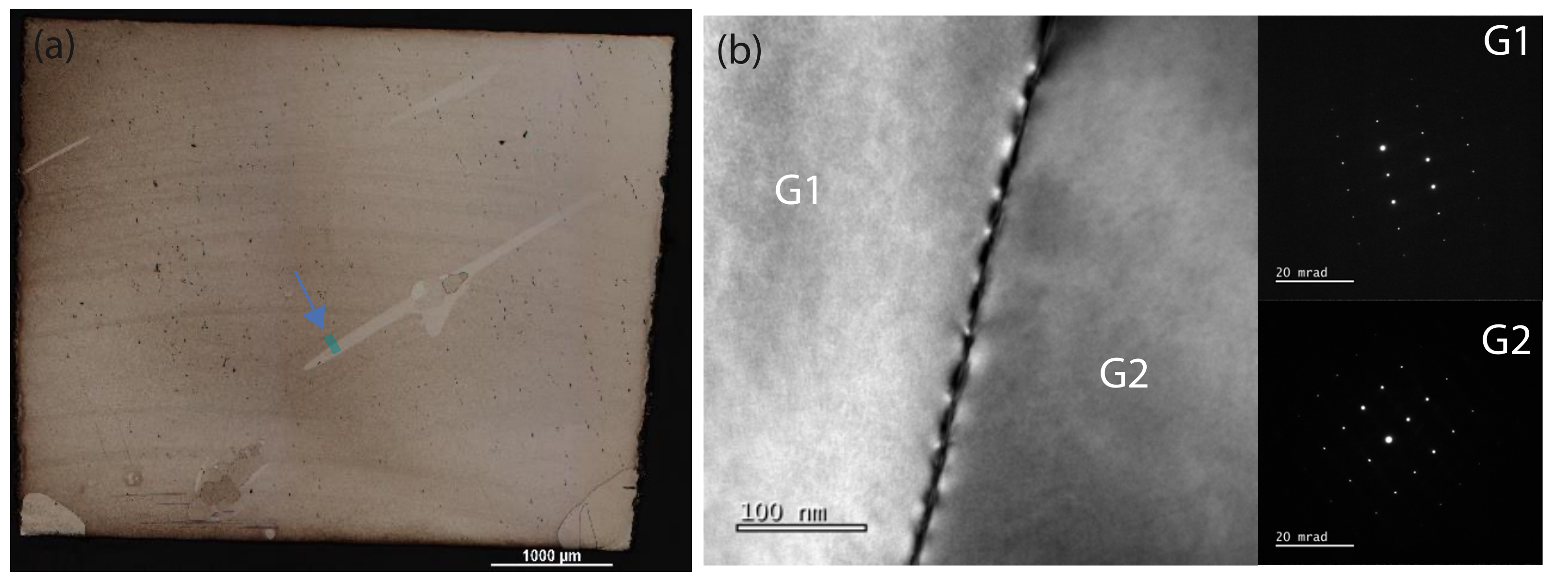}
\caption{(a) Optical micrograph of a compressed \hkl(101) oriented single crystal of $\beta$-Sn with the approximate FIB lift-out region spanning a \hkl(301) twin shown in blue. (b) Bright-field TEM of the twin boundary, which exhibits defects indicated by strain contrast with an average spacing of 30 nm. Diffraction patterns are shown for each grain.}
\label{fig:ebsd}
\end{figure}

A single crystal of nominally pure $\beta$-Sn with a \hkl(101)-orientation obtained from Princeton Scientific was quasi-statically compressed at a strain rate of 0.001 $s^{-1}$ at room temperature (294 K) to a true strain of 0.25. The deformed single crystal was sectioned to reveal a microstructure containing several deformation twins (Fig. \ref{fig:ebsd}a). To further probe the structure of the largest observed deformation twin, a Transmission Electron Microscopy (TEM) lamella was lifted out using a Helios 600 dual Focused Ion Beam (FIB)/ Scanning Electron Microscope (SEM), with a final polishing step at 2 kV to reduce amorphous FIB damage. TEM was conducted at 300 kV on a FEI Titan 80-300 (S)TEM. Transmission Kikuchi Diffraction (TKD) analysis was performed on the lamella inside an FEI Apreo SEM with an accelerating voltage of 30 kV. TKD analysis indicates a predominant misorientation of \hkl[010] 62$^\circ$, consistent with a \hkl(301) twin. TEM diffraction patterns (Fig. \ref{fig:ebsd}b) and TKD results indicate that the twinned grain is well-aligned with the \hkl[010] direction, the tilt axis of the twin boundary. 

The structure of the twin boundary in Fig. \ref{fig:ebsd}b will be shown at higher resolution and compared to simulation results from atomistic simulations in Section \ref{sec:discussion}. 

\subsection{Interatomic potentials}
{\emergencystretch 2em

In this work, two types of ML potentials were developed for $\beta$-Sn: a Moment Tensor Potential (MTP) and a Rapid Artificial Neural Network (RANN) potential. Twinning behavior is compared across these potentials and a classical MEAM potential \cite{Ko:2018}. 

An MTP is characterized by the expansion of local atomic 
environments into symmetry-preserving ``moment tensors'' that capture geometric 
features such as distances and angles \cite{MTP}. By combining these moment tensors with fitted 
coefficients, the potential energy surface of a system can be described in a flexible 
manner. For this study, the training of the MTP was performed using the \texttt{MLIP} \cite{novikov2020mlip} package, 
which offers efficient parameter optimization and convenient integration with 
LAMMPS \cite{LAMMPS}. The MTP is capable of fitting both energies and forces, thus reducing the size of 
the required training set compared to models that fit only energies.  3615 DFT configurations were used for training.

The RANN potential represents an artificial neural network based approach \cite{dickel2021lammps} designed 
to approximate a material’s potential energy surface while remaining computationally 
efficient. Local atomic environments are transformed into numerical descriptors, 
and a feed-forward neural network then produces atomic energies whose sum yields the 
total energy of the configuration; forces are obtained by differentiating these 
energies. Unlike MTP, RANN fits only energies, which necessitates a larger number 
of reference data points to achieve adequate precision on both energies and forces. Approximately 20,007 DFT configurations were 
used in training. Training of the RANN potential in this work was carried out using the \texttt{RANN} \cite{Nitol:2023} calibration tool. Further details of the training procedure and a description of the DFT training database for both the MTP and RANN potentials are given in the Supplementary Information.   
}

The RANN potential developed in this work is a second generation version of the RANN potential published by Nitol \textit{et al.} \cite{Nitol:2023}. This second generation RANN potential improves the thermal stability of GBs and dislocations in $\beta$-Sn compared to the original version. A spurious phase transition to a simple cubic-like phase was observed to heterogeneously nucleate at GBs and dislocations with the original potential. Such a phase is not present in the phase diagram of Sn. Improved stability of the second generation potential was achieved by adding simple cubic structures with varying perturbations to the DFT training database, thereby raising the energy of the simple cubic phase and preventing its formation.


\subsection{Twin boundary structure optimization and migration}

Relaxed GB structures were generated via a grand canonical optimization routine implemented in the open source GRIP tool \cite{GRIP}, which samples a more diverse set of GB structures than the traditionally used $\gamma$-search method \cite{olmsted2009survey} which only samples in-plane translations. A variety of $\hkl<010>$ symmetric tilt GB structures were optimized using the MTP potential, whereas only the \hkl(301) and \hkl(101) twin boundary structures were optimized using RANN and MEAM potentials. Minimum energy structures were used for subsequent migration simulations. 

The GRIP algorithm samples the full range of average atomic densities $[n]$ at each symmetric GB bounded by \hkl{h k l} planes by deleting atoms of one crystal within a slice of width $d_{hkl}$ adjacent to the GB plane, where

\begin{equation}
d_{hkl} = \frac{1}{\sqrt{\frac{h^2+k^2}{a^2} + \frac{l^2}{c^2}}} \, ,
\end{equation}
and 
\begin{equation}
[n] = \frac{N_{tot} \ \text{mod}(N_{hkl})}{N_{hkl}} \, .
\end{equation}

\noindent $N_{tot}$ is the total number of atoms in the system and $N_{hkl}$ is the total number of atoms within a slice of width $d_{hkl}$ such that $[n]$ lies in the range 0 to 1. Whereas the commonly employed $\gamma$-search method makes the assumption that minimum energy structures will be found at $[n] = 0$ or equivalently $[n] = 1$, the GRIP method allows the discovery of additional low energy structures at intermediate $[n]$ \cite{hickman2017extra, zhu2018predicting}.  In addition to changes in density, relative translations of the grains are sampled at random in the GRIP method followed by an annealing and quenching protocol. A full discussion of structural perturbations employed in the GRIP method is found in \cite{GRIP}. 

The GRIP method was employed in \textit{user} mode with atomic coordinates for each grain input by the user. Atom coordinates of the upper and lower grains were generated using the Atomic Simulation Environment (ASE \cite{ase-paper}) surface class with surface indices $\hkl(h 0 l)$ and surface normal parallel to the sample $Z$ direction. A $\hkl[100] 180 ^\circ$ twinning rotation was applied to the lower grain such that the sample directions were $X_{lower} \parallel \hkl[-l 0 h]$, $Y_{lower} \parallel \hkl[0 -1 0]$ and $X_{upper} \parallel \hkl[l 0 -h]$, $Y_{upper} \parallel \hkl[0 1 0]$. For all simulations, periodic boundaries were employed along the in-plane $X$ and $Y$ directions, but not along the $Z$ direction normal to the interface.

A different philosophy was pursued for creation of twin inclusion structures. Metastable states were generated without significant effort devoted to structure optimization. Twin boundaries are expected to find low energy structures during migration and additional information about growth mechanisms can be obtained via the pinning effect of metastable GB structures. Twin nucleation mechanisms in $\beta$-Sn are not currently well understood \cite{tu1970direct} and a thorough survey of heterogeneous nucleation mechanisms is out of the scope of the current work. Metastable twin boundary structures were created by rotating a cylindrical inclusion with a specified misorientation and radius within an initially single crystal. For pairs of atoms with pairwise distance less than 2 {\AA}, one atom was deleted at random. Finally, a conjugate gradient minimization was performed with energy and force tolerances $10^{-8}$ eV and $10^{-8}$ eV/{\AA}. 

To simulate GB migration and twin growth, a pure shear strain $\epsilon_{xz}$ was applied to each system by displacing fixed slabs at the top and bottom of the system in opposite directions perpendicular to the tilt axis. We denote the sample $X$ direction as the shear direction. Various temperatures were simulated in the iso-thermal NVT ensemble with structures uniformly expanded by a pre-computed thermal expansion coefficient prior to loading. A strain rate of $\sim 1 \times 10^8\ $ s$^{-1}$ was employed with a fixed slab thickness of 0.5 nm. Strain control was employed, rather than stress control, for convenience in analyzing stick-slip behavior. The shear coupling factor was measured as the total shear distance divided by the total migration distance and was averaged over two independent simulations with opposite signs of applied shear strain. A negative coupling factor corresponds to downward GB migration with a positive applied shear strain or upward GB migration with a negative applied shear strain. 

System size is an important consideration in simulations of GB migration \cite{race2014role}. For measurements of shear coupling factor, quasi-2D bicrystal systems employed a simulation box with length of approximately 18 nm along the shear direction which was sufficient to observe disconnection nucleation events without rapid annihilation across periodic boundaries. Simulation length along the tilt axis was 1.5 nm. Grain sizes in the interface plane normal direction were 25 nm for a total simulation box height of 50 nm. In 3D simulations with flat bi-crystals, larger systems were considered for the \hkl(301) CTB versus \hkl(101) CTB because of the lower critical shear stress of the former for migration which enabled faster simulations of migration. For \hkl(301) CTB simulations with all potentials and \hkl(101) CTB simulations with the MEAM potential, GB plane sizes of 18 nm x 18 nm were considered. For \hkl(101) CTB simulations with the ML potentials, GB plane sizes of 9 nm x 9 nm were considered. All twin growth simulations with inclusions were quasi-2D with dimensions of approximately 40 nm in the shear and interface plane normal direction and 1.5 nm along the tilt axis. 

\section{Results}
\label{sec:results}

\subsection{The structure and energy of coherent twin boundaries}

\begin{figure}[H]
\centering\leavevmode \includegraphics[width=0.99\textwidth]{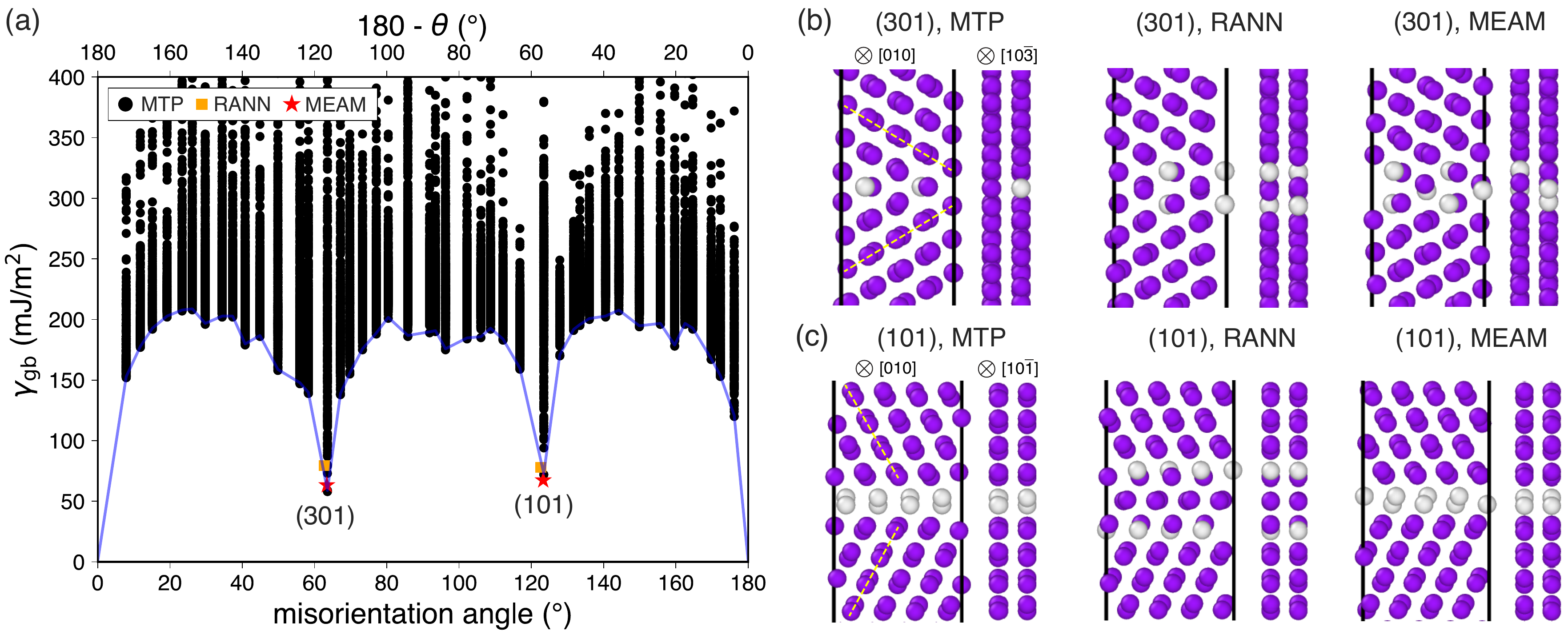}
\caption{Energy and structure of twin boundaries in $\beta$-Sn across potentials (a) The GB energy landscape for \hkl[010] symmetric tilt boundaries as computed with the MTP exhibits cusps corresponding to the \hkl(301) and \hkl(101) twins. Minimum energies of these twins are also shown for the RANN and MEAM potentials. (b-c) Minimum energy structures of the \hkl(301) and \hkl(101) twin boundaries vary slightly across potentials. Front and side views are given for each twin. Dashed lines indicate \hkl(-101) type planes separated by the misorientation angle.}
\label{fig:energy}
\end{figure}

The \hkl(301) and \hkl(101) twin boundaries have the lowest energies among high angle \hkl[010] symmetric tilt GBs in $\beta$-Sn as shown in Fig. \ref{fig:energy}a. These twin boundaries are associated with deep local minima in the GB energy landscape. Twin boundary energies are given in Table \ref{tab:twin} and span a range of 57-79 mJ/m$^2$, less than half of the average minimum energy of high angle GBs in $\beta$-Sn, 185 mJ/m$^2$. The spectrum of energy values obtained by the MTP simulations including metastable states 57-450 mJ/m$^2$, compares well to experimental values spanning 140-600 mJ/m$^2$ \cite{mykura1955, rowenhorst2005measurements, ware2022}. The experimental values are indirect measurements which average over multiple types of GBs and rely on reference values for surface energy \cite{rowenhorst2005measurements, ware2022}. To the authors' knowledge, an experimental estimate for the energies of the specific \hkl(301) or \hkl(101) twin boundaries has not yet been obtained.

The ordering of \hkl(301) and \hkl(101) twin boundary energies is found to vary with interatomic potential as shown in Fig. \ref{fig:energy} and Table \ref{tab:twin}. The MEAM potential predicts the \hkl(101) twin as having a lower energy than the \hkl(301) twin by 2 mJ/m$^2$whereas the ML-based RANN potential and MTP predict a lower energy for the \hkl(301) versus \hkl(101) twin by 10-15 mJ/m$^2$. DFT calculations were attempted to determine the relative ordering of GB energies at 0 K but were unable to converge with input structures containing over 100 atoms each for the \hkl(301) and \hkl(101) twins. The only prior calculation of twin boundary energy in $\beta$-Sn that we are aware of is a DFT study of Cu diffusion near the \hkl(101) CTB in $\beta$-Sn \cite{hao2023effect}, which found a large energy of 630 mJ/m$^2$ beyond the experimentally measured spectrum of values. Such a large value calls into question the accuracy of the results reported in \cite{hao2023effect}. We note that $\beta$-Sn is metastable at 0 K and that a full assessment of relative GB energies would require free energy calculations at finite temperatures \cite{freitas2018free}. 

The twin boundaries exhibit compact core structures spanning 2-4 layers of disordered atoms with slight differences in the positions of atoms in the core region depending on the interatomic potential (Fig. \ref{fig:energy}b). Minimum energy structures were found in all cases with $[n] = 0$, implying that the standard $\gamma$-search method is sufficient for optimization of these CTBs. This is contrast to GBs with misorientations near the CTBs shown in Figs. \ref{fig:near101} and \ref{fig:near301}, which frequently have minimum energy structures with $[n] \ne 0$. The differences in CTB structure across potentials can be rationalized by the multi-lattice structure of $\beta$-Sn. The GB can be centered at one of two $(301)$ or $(101)$ planes in the vertical direction with pairs of atoms in the GB core rotating into horizontal or vertical conformations depending on the choice of plane. The minimum energy MTP \hkl(301) CTB structure is centered at a different $(301)$ plane compared to the RANN or MEAM structures. The RANN \hkl(101) CTB is centered on a different \hkl(101) plane than the MTP or MEAM structures. Structures are found to have a small microscopic shift in the range of 0.1-0.6 \AA\ along the shear direction (horizontal axis in Fig. \ref{fig:energy}b) with minimal translation along the tilt axis. HRTEM images of the \hkl(101) CTB obtained in this work will be discussed in Section \ref{sec:twin_growth}. 

\begin{table}[H]
    \centering
    \begin{tabular}{lccc c}
        \hline
        Property & MEAM & RANN & MTP & Experiment / Theory \\
        \hline
        $c/a$ & 0.547 & 0.540 & 0.538 & 0.546 \cite{Nitol:2023} \\
        $\theta_{301} (^\circ)$ & 62.7 & 63.4 & 63.6 & 62.8 \\
        $\theta_{101} (^\circ)$ & 122.6 & 123.3 & 123.5 & 122.7 \\
        $\gamma_{301}$ (mJ/m$^2$) & 79.1 & 58.9 & 57.3 & - \\
        $\gamma_{101}$ (mJ/m$^2$) & 77.9 & 68.2 & 72.4 & - \\
        $\gamma_{AP, 1:2}$ (mJ/m$^2$) & 88 & 55.1 & 71.2 & - \\
        $\gamma_{\text{general}}$ (mJ/m$^2$) & - & - & 185 $\pm$ 19 & 160 $\pm$ 40 \cite{mykura1955}, \\
        &&&& 200-600 \cite{ware2022} \\
        $\beta_{301}$ & -0.109 (-0.089)  & -0.096 (-0.122) & -0.101 (-0.121) & -0.0982 \cite{molodov2018grain}, \\
        &&&& $|\beta|$ = 0.0978 \cite{tu1970direct}, 0.113\cite{christian2002} \\
        $\beta_{101}$ & 1.07 (1.23) & -0.100 (-0.115) & -0.103 (-0.121) & -0.0982  \cite{molodov2018grain}, \\
        &&&& $|\beta|$ = 0.0978 \cite{tu1970direct}, 0.113\cite{christian2002}\\
        \hline
    \end{tabular}
    \caption{Twin boundary properties across potentials in comparison to reference values from experiment or theory: $c/a$ ratio, CTB misorientation angles, interface energies ($\gamma$) for CTBs, coherent AP facets and general boundaries. Shear coupling factors ($\beta$) for CTBs with simulation values at 300 K shown with 0 K values in parentheses.}
    \label{tab:twin}
\end{table}

\subsection{Shear coupled migration of coherent twin boundaries}
\label{sec:sc}

\begin{figure}[H]
\centering\leavevmode \includegraphics[width=0.99\textwidth]{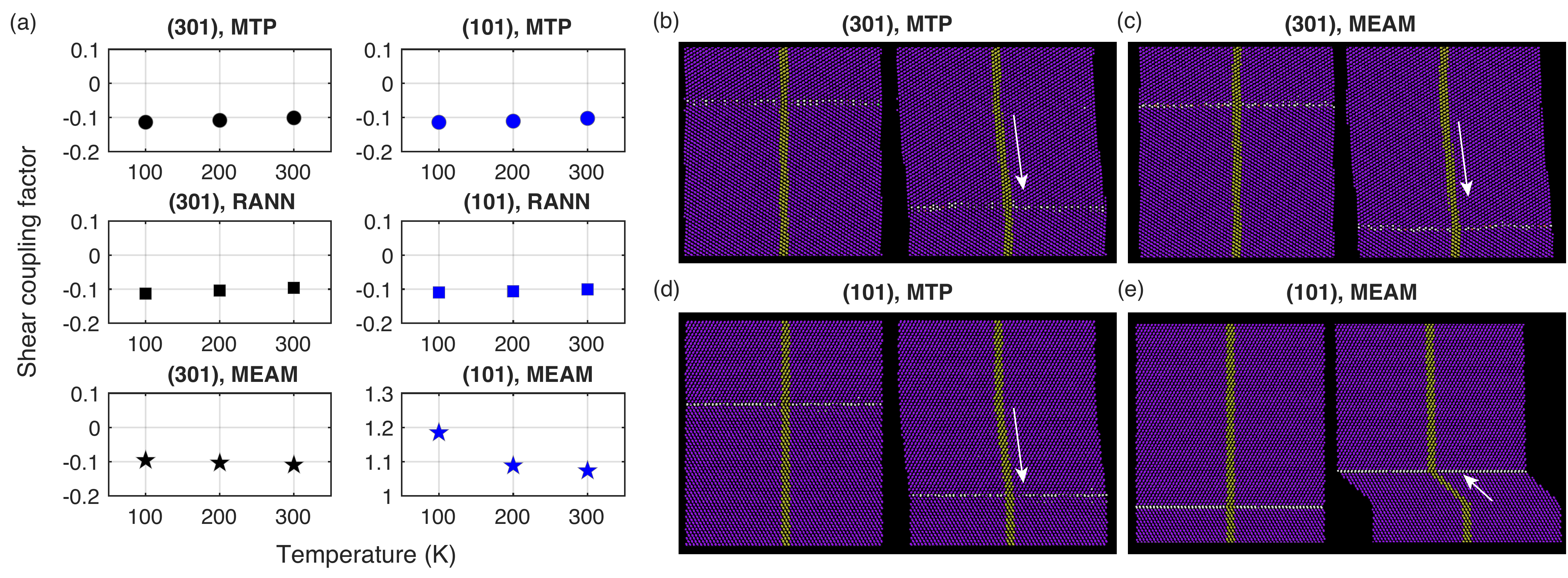}
\caption{Shear coupling behavior of twin boundaries in $\beta$-Sn across potentials (a) Shear coupling factor as a function of temperature. (b)-(e) Snapshots before and after GB migration for the MTP and MEAM potentials at 100 K with fiducial markers shown in yellow and the direction of GB migration indicated by arrows. Bicrystal dimensions are 18 nm along the horizontal \hkl[-l 0 h] direction, 50 nm in the vertical direction and 1.5 nm into the page along the \hkl[010] tilt axis.}
\label{fig:sc}
\end{figure}

Shear coupling factors measured in quasi-2D twin boundary migration simulations (Fig. \ref{fig:sc} and Table \ref{tab:twin}) support the prior hypotheses \cite{tu1970direct, molodov2018grain} that the coupling factors for \hkl(301) and \hkl(101) CTBs are the same. Small and negative coupling factors of around -0.1 are measured for all interatomic potentials and temperatures except for the \hkl(101) twin with the MEAM potential, which exhibits a large positive coupling factor around 1.2 which decreases with increasing temperature. Snapshots before and after shear coupled twin boundary migration are shown in Fig. \ref{fig:sc}b-e for the MTP and MEAM potentials. The \hkl(101) MEAM twin boundary moves in the opposite direction of all the other boundaries for the same direction of applied shear strain, implying a different migration mechanism with a larger degree of GB sliding compared to the \hkl(101) migration simulations using the MTP.

\begin{figure}[H]
\centering\leavevmode \includegraphics[width=0.5\textwidth]{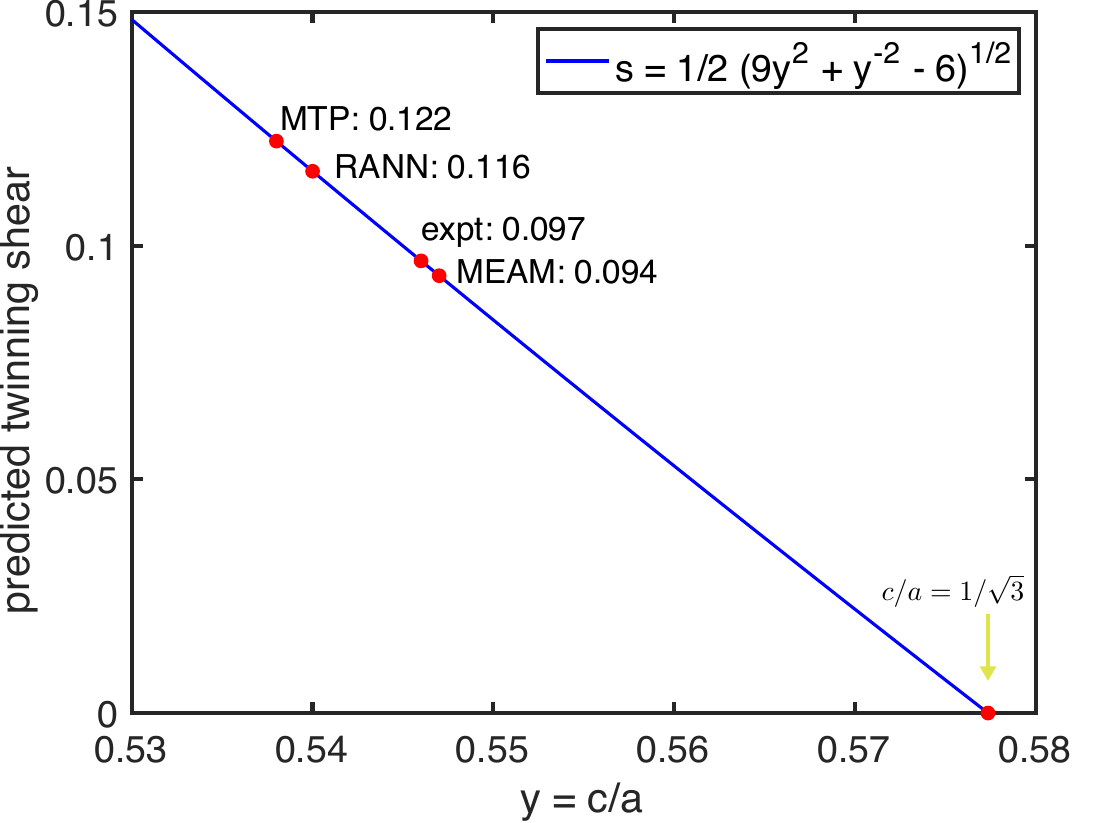}
\caption{Twinning shear as a function of c/a ratio \cite{christian2002}  for ratios given in Table \ref{tab:twin}.}
\label{fig:ca}
\end{figure}

Christian \textit{et al.} \cite{christian2002} gave a $c/a$ ratio dependent formula for the twinning shear in $\beta$-Sn

\begin{equation}
|\beta| = 1/2 \sqrt{9 \gamma^2 + \gamma^{-2} - 6} \, ,
\label{eqn:ca}
\end{equation}

\noindent with $\gamma = c/a$. A linear decrease in twinning shear magnitude is predicted with increasing $c/a$ ratio as shown in Fig. \ref{fig:ca}. This trend matches the MD data well with less than 5\% variation between theory and simulation for 0 K data. This comparison ignores the outlying data for the MEAM \hkl(101) CTB, which illustrates that multiple mechanisms with distinct coupling factors / twinning shears are possible to accommodate CTB migration. GBs support a countably infinite number of disconnection modes, only one or a few of which are accessible in the energy landscape \cite{han2018grain, chesser2021optimal, rohrer2023grain, joshi2024}. To the best of our knowledge, shear coupling factors for CTBs in $\beta$-Sn have never directly been measured experimentally. Coupling factors have been inferred from electron diffraction data with the assumption of minimum twinning shear \cite{tu1970direct} or geometrically enumerated \cite{molodov2018grain}. 

\begin{figure}[H]
\centering\leavevmode \includegraphics[width=0.99\textwidth]{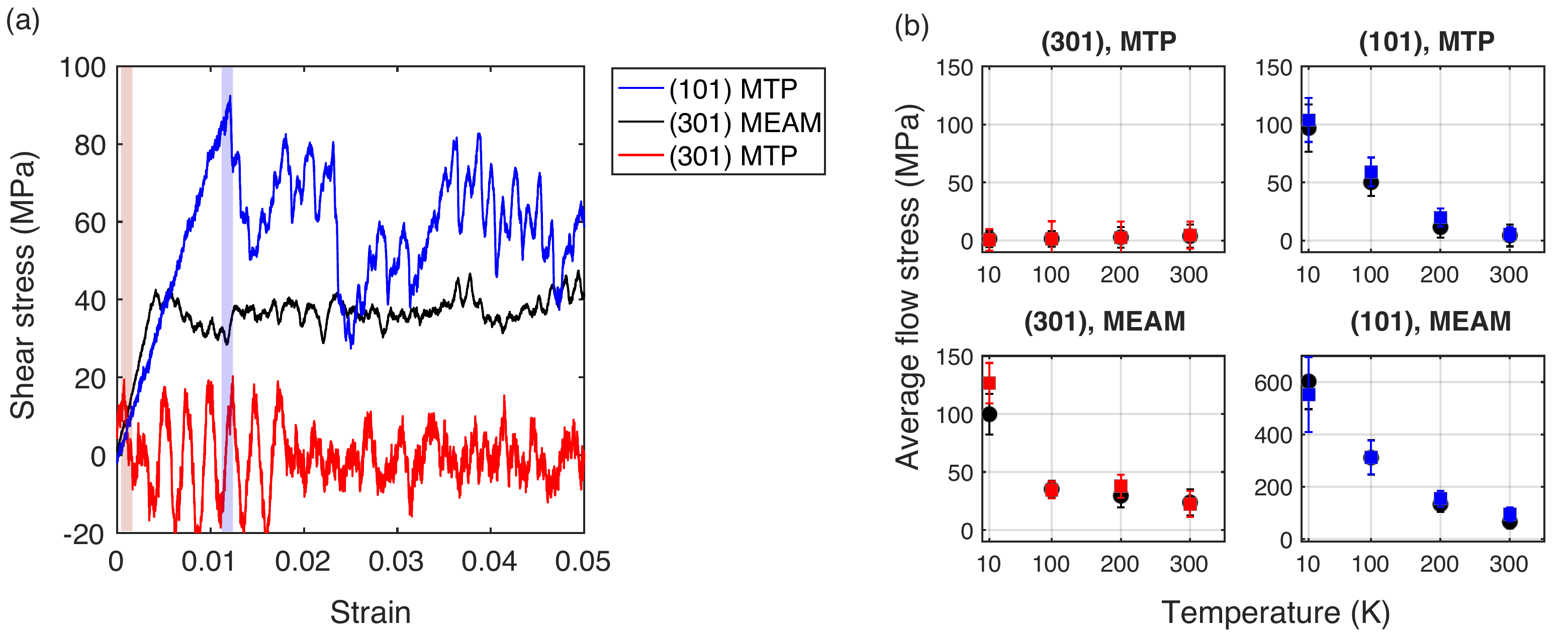}
\caption{Stress-strain behavior for CTB migration (a) Stress-strain curves for selected boundaries and potentials at 100 K. Highlighted strain intervals correspond to the onset of net twin boundary migration for the two twin types using the MTP potential. (b) Average flow stress as a function of temperature for the MTP and MEAM potentials. Black points are quasi-2D simulations, whereas colored points are fully 3D simulations. }
\label{fig:ss}
\end{figure}

\begin{figure}[H]
\centering\leavevmode \includegraphics[width=0.65\textwidth]{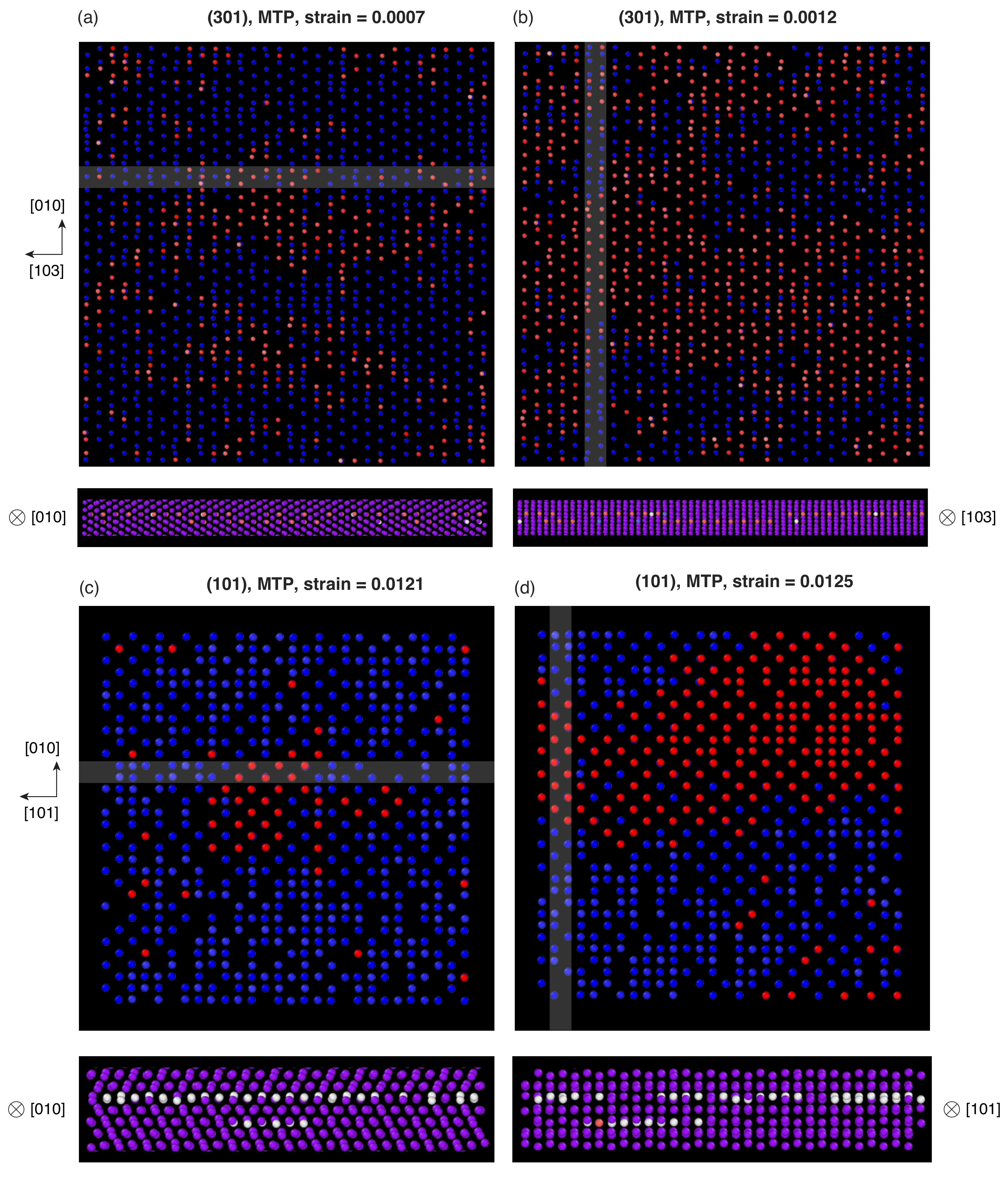}
\caption{Disconnection island nucleation geometry for the (a)-(b) \hkl(301) and (c)-(d) \hkl(101) CTBs at the onset of GB migration (see highlighted strain intervals in Fig. \ref{fig:ss}) for 3D MTP simulations at 100 K. In each subfigure, the top panel shows a bottom-up view of the twin boundary (migrating downward) with only disordered atoms shown, colored by GB normal coordinate (red and blue represent different layers). The bottom panel shows an edge-on view of the twin with all atoms shown, colored by local environment. Only a slice of the entire structure is shown as given by the region highlighted in grey in the upper panel. In (a)-(b), a PTM RMSD cutoff of 0.4 is used to more clearly highlight the disconnections, whereas a cutoff of 0.25 is used for (c)-(d).}
\label{fig:nucleation}
\end{figure}

Selected stress-strain curves for shear coupled migration are shown in Fig. \ref{fig:ss}a along with flow stress trends as a function of temperature for both quasi-2D (black points) and fully 3D simulations (colored points) in Fig. \ref{fig:ss}b. Stick-slip behavior is found to correspond to disconnection-mediated shear coupled migration \cite{cahn2006coupling}. For the MTP simulations, highlighted stress drops of the stress-strain curves for the \hkl(301) and \hkl(101) twin boundaries in Fig. \ref{fig:ss}a  correspond to the disconnection nucleation events in Fig. \ref{fig:nucleation}. Islands of double layer disconnections are observed to nucleate and grow with step heights $h$ equal to $2 d_{301} = 0.34$ nm and $2 d_{101} = 0.57$ nm. Given the measured shear coupling factors, we predict small Burgers vector magnitudes $\beta h$ of $b_{301} = 0.04$ nm and $b_{101} = 0.06$ nm. We refer to these double layer disconnections as h(2) disconnections following the notation of Hirth et al. \cite{hirth2016disconnections} where a h(n) step has a total height n$d_{h0l}$. Simulations of both twin boundaries are large enough to capture a multiple-island nucleation regime in which multiple nuclei may grow and compete \cite{race2014role, spearot2019shear}. Flow stress values are similar between quasi-2D and 3D systems as shown in Fig. \ref{fig:ss}b, implying that there is not a strong dependence of flow stress on system size along the tilt axis for the GB lengths and strain rates used in this work. 

The \hkl(101) twin boundary exhibits a larger yield strain and flow stress for migration than the \hkl(301) twin boundary by at least an order of magnitude, with yield strains of 0.01 versus 0.001 and flow stresses of 50 MPa versus 0-5 MPa for simulations employing the MTP. The yield strains and stresses are so small for the MTP \hkl(301) twin boundary that they are difficult to distinguish from noise in the overall stress signal due to the thermostat. The flow stress is assumed to lie in the range between 0 MPa and the maximum observed stress of 5 MPa. Whereas the \hkl(101) twin boundary exhibits a decrease of flow stress with increasing temperature, the \hkl (301) twin boundary exhibits athermal behavior with little variation of flow stress with temperature. Near room temperature at 300 K, both boundaries exhibit similarly small flow stresses. The athermal behavior of the \hkl(301) MTP boundary contrasts with the \hkl(301) MEAM boundary that exhibits a flow stress of around 100 MPa at 10 K, which decreases with increasing temperature. The \hkl(101) MEAM boundary exhibits a large flow stress of 600 MPa at 10 K with a decrease in flow stress with increasing temperature. Such a large flow stress can be attributed to a disconnection mode with a large Burgers vector and small step height identified in Table \ref{tab:bv} and shown in Fig. \ref{fig:CSL}c. 

\subsection{Twin growth and bicrystallography}
\label{sec:twin_growth}

\begin{figure}[H]
\centering\leavevmode \includegraphics[width=0.99\textwidth]{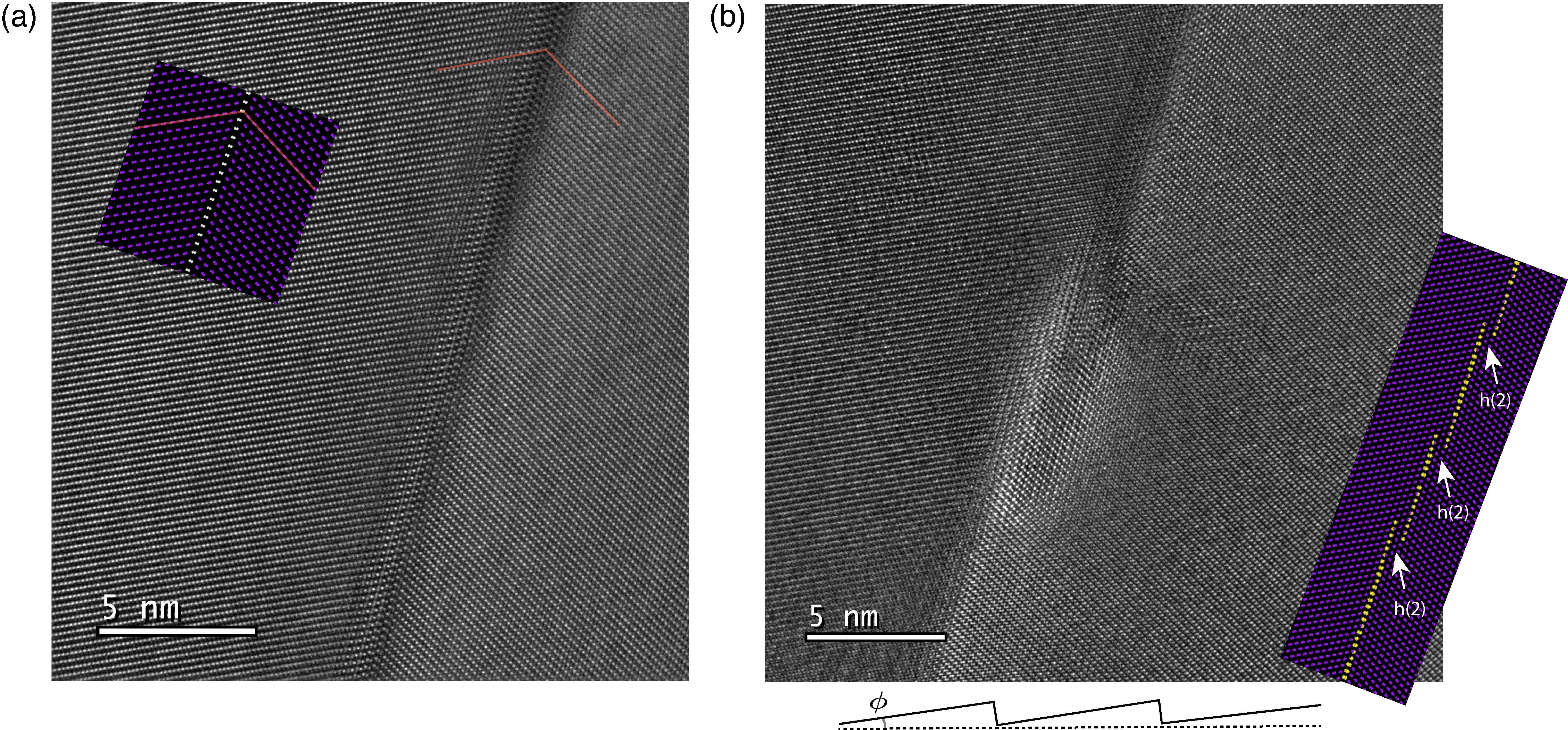}
\caption{HRTEM images obtained in this work of a semi-coherent boundary delineating a \hkl(301) twin with (a) \hkl(-101) CTB segments separated by (b) defective regions with both step and dislocation character. The angular marker in (a) is calibrated to MTP simulation data and superimposed on the experimental data, showing a slight deviation in apparent tilt angle in the experiment versus simulation. An observed growth morphology from simulation is also compared to the experimental data in (b) with unit h(2) twinning disconnections identified.}
\label{fig:exptCTB}
\end{figure}

To validate the predictions of the simulations in this work, we compare twin boundary structures from MTP simulations to experimentally observed twin boundary structures shown in Fig. \ref{fig:ebsd} and Fig. \ref{fig:exptCTB}. Fig. \ref{fig:exptCTB} shows HRTEM images of a twin with a \hkl(301) type misorientation at different points along its length including a coherent region (Fig. \ref{fig:exptCTB}a) and a defective region with a step (Fig. \ref{fig:exptCTB}b). The observed growth morphology is surprising in the sense that the growth plane along the long axis of the twin is not the low energy \hkl(301) CTB, but rather a semi-coherent boundary with segments of near-\hkl(101) CTB separated by defects with both step and dislocation character. Measurements of misorientation angle from 80 distinct pairs of \hkl{101} planes along the coherent interface in Fig. \ref{fig:exptCTB}a give an angle $125.6\pm0.7^\circ$ consistent with a near-\hkl(-101) CTB rather than the \hkl(301) CTB. 

\begin{figure}[H]
\centering\leavevmode \includegraphics[width=0.9\textwidth]{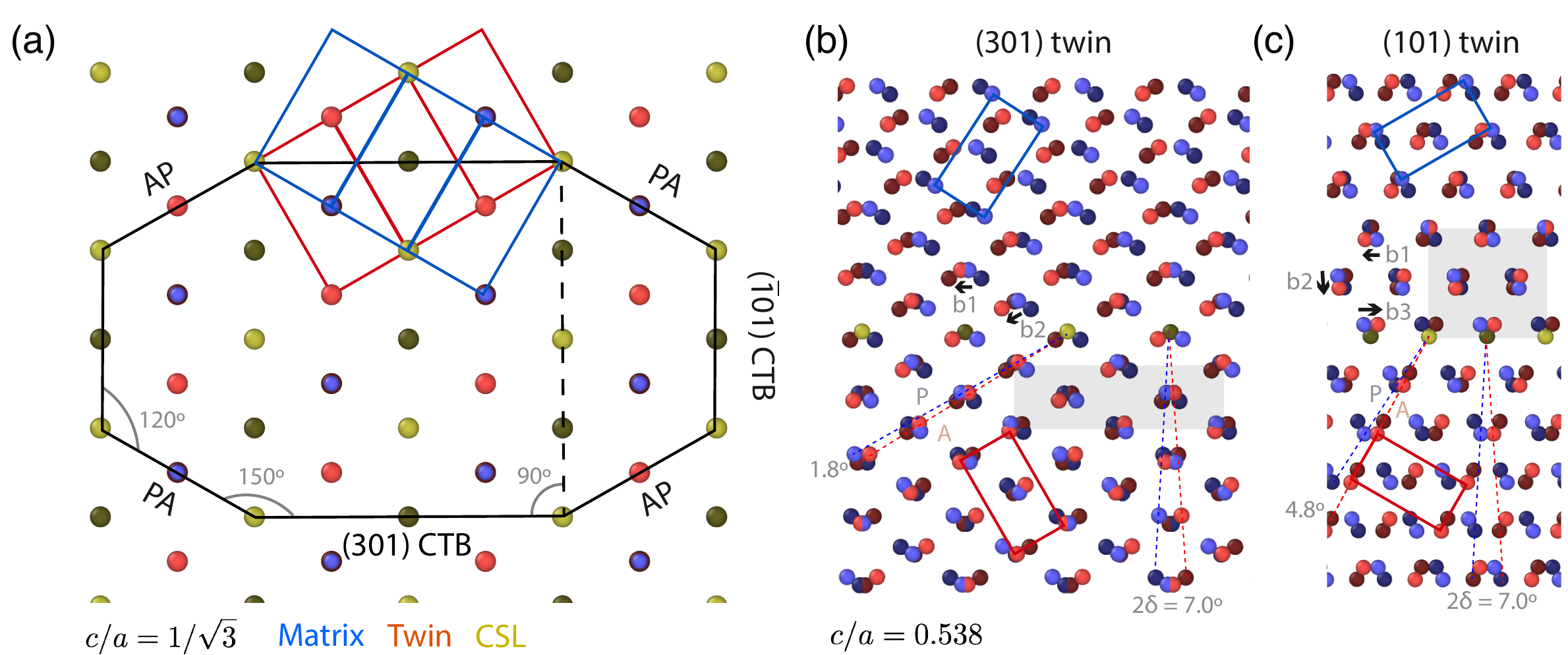}
\caption{(a) A reference twin with ideal c/a ratio is shown with a \hkl[010] 60$^\circ$ misorientation corresponding to a $\Sigma2$ dichromatic pattern. A twin inclusion is outlined by the solid black lines with interface types and angles labeled. Only BCT lattice sites are shown. The dichromatic patterns for twins in actual $\beta$-Sn are given for the (b) \hkl(301) twin and (c) \hkl(101) twin with MTP lattice parameters. Atoms in the matrix grain (blue), twin grain (red) and at CSL points (yellow) are shown in light or dark contrast corresponding to the front and back planes along the tilt axis (into the page). In (b)-(c), the angular deviations from ideal character of the 90$^\circ$ conjugate twin facets and AP/PA interfaces are highlighted. Burgers vectors $b_i$ for all disconnection modes observed in this work are labeled and enumerated in Table \ref{tab:bv}. $b_1$ in each figure corresponds to the unit h(2) twinning disconnection which mediates CTB migration. Grey boxes correspond to unit cells for which shuffling patterns are shown in Fig. \ref{fig:shuffle}.}
\label{fig:CSL}
\end{figure}

The observed growth morphology of the \hkl(301) twin can be understood from bicrystallography and a loading state that favors growth of facets 90$^\circ$ from the \hkl(301) CTB rather than the CTB itself. A high symmetry reference bicrystal with ideal c/a ratio equal to $1/\sqrt{3} = 0.577$ is shown in Fig. \ref{fig:CSL}a. This ideal c/a ratio is larger than the actual c/a ratio of $\beta$-Sn, which is close to $\sqrt{3/10} = 0.548$. In this hypothetical crystal, a twin inclusion with a misorientation \hkl[010] 60$^\circ$ has a $\Sigma 2$ coincident site lattice (CSL) with \hkl(301) and \hkl(-101) CTB facets which lie exactly 90$^\circ$ apart. The predicted twinning shear is zero as shown in Fig. \ref{fig:ca}.  The nonideal c/a ratio of actual $\beta$-Sn breaks the symmetry of the $\Sigma 2$ dichromatic pattern in a way that leads to 1) distinct misorientations for each CTB, 2) nonzero twinning shears/coupling factors for CTBs associated with disconnections with finite Burgers vectors and 3) an additional disclination character of non-CTB facets superimposed upon their ideal inclinations in the $\Sigma2$ dichromatic pattern. In $\beta$-Sn, \hkl(301) and \hkl(-101) type planes lie $(90-\delta)^\circ$ apart, where $\delta$ is defined as the obliquity parameter \cite{grimmer2003,grimmer2004} such that


\begin{equation}
2\delta = \tan^{-1}|\beta| \, .
\label{eqn:tan}
\end{equation}

\noindent The misorientation angles of (301) and (-101) CTBs are then expressed as approximately ($60+\delta$) and ($180-(60-\delta)$). Twin facets which are observed to lie 90$^\circ$ from (301) or (101) CTBs have disclination character 2$\delta$ superimposed upon \hkl(-101) or \hkl(-301) CTBs. The oiLAB software package \cite{oiLAB} is used to further characterize the bicrystallography of twin boundaries for each twin type in $\beta$-Sn including $\Sigma$ values, non-ideal facet inclination angles and disconnection modes as discussed in the Supplementary Information.

\begin{figure}[H]
\centering\leavevmode \includegraphics[width=0.95\textwidth]{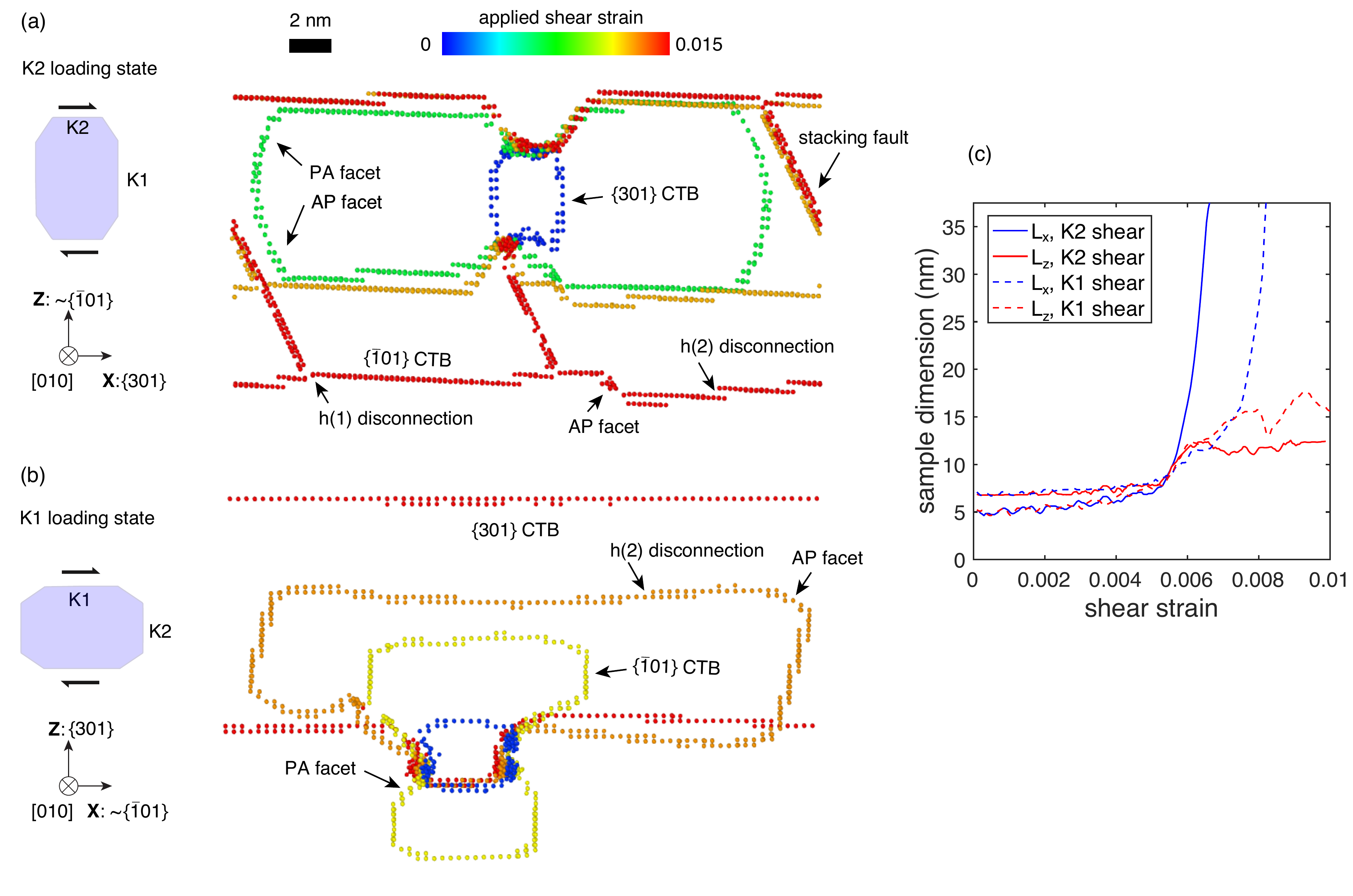}
\caption{Twin growth trajectories for \hkl(301) inclusions at 10 K depend on loading geometry. (a) shows a K1 loading state with shear resolved along the \hkl(301) CTB, whereas (b) shows a K2 loading state with shear resolved along the 90$^\circ$ boundary close to the \hkl(-101) CTB. Outlines of the twin inclusions with atoms identified as disordered are shown at different times corresponding to increasing applied shear strain. Features of twin growth are labeled including facet types and observed defects. (c) Illustrates changes in twin dimensions during growth along the shear (X) and normal (Z) directions.}
\label{fig:tgmtp}
\end{figure}

Growth morphologies of \hkl(301) twins from direct MD simulations using the MTP are shown in Fig. \ref{fig:tgmtp} for two loading conditions. In Fig. \ref{fig:tgmtp}a, a shear strain is resolved upon the facet 90$^\circ$ from the \hkl(301) CTB, whereas Fig. \ref{fig:tgmtp}b shows the growth of the same twin inclusion with a shear strain resolved upon the \hkl(301) CTB. As a shorthand notation, we denote the first loading state as \textit{K2 shear} and the second loading state as \textit{K1 shear}. The \textit{K2 shear} loading state produces a semi-coherent interface structure consistent with the experimentally observed growth morphology. 

For both loading states, the twin inclusion grows rapidly in the shear direction (sample X direction) and more slowly in the normal direction (sample Z direction) as shown in Fig. \ref{fig:tgmtp}c. Twin tips are observed to annihilate across periodic boundary conditions, sometimes leaving a stacking fault (SF) behind as in Fig. \ref{fig:tgmtp}a. After annihilation, isolated twin facets move independently to thicken the twin. For \textit{K1 shear}, the uppermost \hkl(301) CTB is able to recover a flat structure after annihilation. For \textit{K2 shear}, a semi-coherent structure is apparent after annihilation with segments of \hkl(-101) CTB separated by disconnections. Two types of disconnections are observed: h(2) disconnections which mediate CTB migration and h(1) disconnections at the intersection of the CTB and SF. These disconnection modes are labeled $b_1$ and $b_2$ in the dichromatic pattern shown in Fig. \ref{fig:CSL}b and are enumerated in Table \ref{tab:bv}. An important difference in the microstructural evolution between the two loading scenarios is that, in the case of \textit{K2 shear}, the entire inclusion rotates by around 2.5$^\circ$ during growth, whereas in the case of \textit{K1 shear} no net rotation is observed. The net rotation in the former scenario results in a semi-coherent boundary that has asymmetric tilt character and a stacking fault which forms at the intersection of twin tips due to incompatibility. 

\begin{table}[ht]
\centering
\begin{tabular}{l l l}
\hline
(301) Twin & \textbf{b} (\AA) & \textbf{h} (\AA)  \\
\hline
$b_1$ & [\,-0.39, 0, 0\,] & $2d_{301} = 3.39$ \\
$b_2$ & [\,-0.86, 0, -0.42\,] & $d_{301} = 1.69$ \\
\hline
(101) Twin & \textbf{b} & \textbf{h} \\
\hline
$b_1$ & [\,-0.65, 0, 0\,] & $2d_{101} = 5.66$ \\
$b_2$ & [\,-0.08, 0, -0.71\,] & $d_{101} = 2.83$ \\
$b_3$ & [\,0.81, 0, 0\,] & 0.71 \\
\hline
\end{tabular}
\caption{Enumeration of disconnection modes using MTP lattice parameters. Corresponding Burgers vectors are shown in Fig. \ref{fig:CSL}b. Unit h(2) disconnections have Burgers vectors labeled $b_1$ for each twin. h(1) disconnections have Burgers vectors labeled $b_2$. The high shear mode observed for the MEAM \hkl(101) CTB is labeled $b_3$.}
\label{tab:bv}
\end{table}

Simulation snapshots of the \hkl(-101) CTB and the semi-coherent facet are compared with the HRTEM images in Fig. \ref{fig:exptCTB}. The angular marker shown in red along the coherent boundary in Fig. \ref{fig:exptCTB}a is calibrated to 0 K MTP simulation data with misorientation $\theta_{101, MTP} = 123.5^\circ$, illustrating a small deviation in angle of around $2^\circ$ between simulation and experiment. The measured angle from the experimental micrograph is also $3^\circ$ larger than the expected value $\theta_{101, expt} = 122.7^\circ$ using a c/a ratio measured in prior experiments \cite{Nitol:2023}. Several factors may contribute to the deviation between measured and predicted CTB angles, including possible facet disclination character, deviations in misorientation angle from the ideal \hkl(301) twin and uncertainty in the temperature dependent c/a ratio measured in prior experiments. 

In the HRTEM image of the defective region in Fig. \ref{fig:exptCTB}b, an estimated total step height (H) of 1.8-2.4 nm is measured corresponding to 6-9$d_{(101)}$. With an average defect spacing of 30 nm (D), the semi-coherent interface is expected to have CTB segments tilted $\phi = \sin^{-1}(H/D) =  3.2\text{-}4.9^\circ$ from the average interface plane. The $\phi$ measured from simulation snapshots for the CTB segments in the boundary in Fig. \ref{fig:exptCTB}b and Fig. \ref{fig:tgmtp}a is approximately 2.6$^\circ$. This corresponds to a geometrically necessary step content of 3$\times$h(2) over the periodic length of the simulation box  in the shear direction (38 nm), motivating a simulated comparison to experiment that contains 3 unit twinning disconnections as shown in Fig. \ref{fig:exptCTB}b. In the experimental structure, multiple unit disconnections appear to coalesce into a relatively narrow 10 nm region along the boundary. Disconnection pile-up structures are observed in our simulations including asymmetric facets discussed in detail below. 

\begin{figure}[H]
\centering\leavevmode \includegraphics[width=0.9\textwidth]{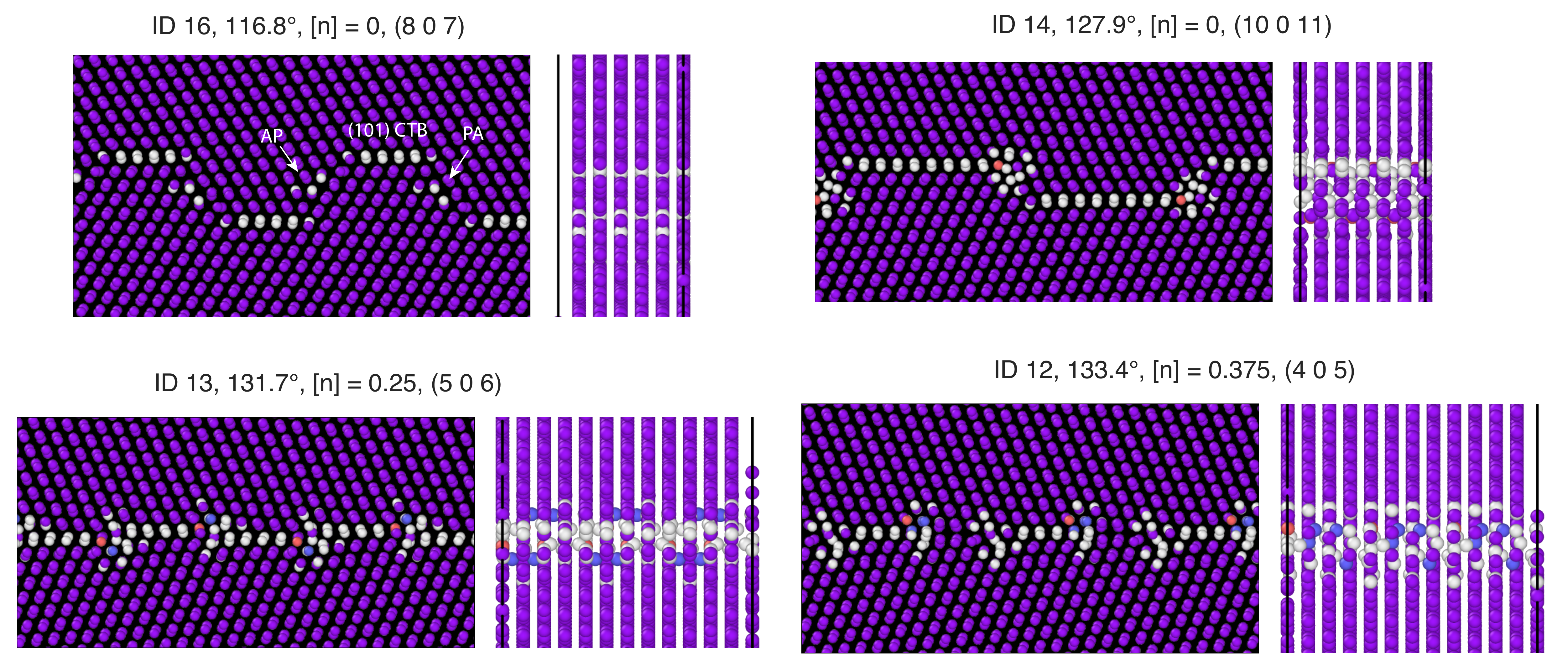}
\caption{Minimum energy symmetric tilt boundary structures with misorientations close to the \hkl(101)  twin ($\theta = 123.5^\circ$) exhibit \hkl(101) CTB structures separated by defective regions. Structures are labeled by ID, misorientation angle, atomic density and boundary plane inclination \hkl(h0l). Snapshots on the left are viewed with the \hkl[010] tilt axis into the page. Snapshots on the right are viewed with the \hkl[-l0h] shear direction into the page. ID 16 corresponds to a GB nearly $90^\circ$ from the \hkl(301) CTB which is observed to facet into AP/PA and \hkl(101) CTB segments.}
\label{fig:near101}
\end{figure}

Fig. \ref{fig:near101} shows predicted structures of twin boundaries with misorientation angles close to the \hkl(101) CTB. Boundaries within $8^\circ$ from the \hkl(101) CTB are found to exhibit patches of coherent CTB-like character separated by facets that contain a combination of step and dislocation character. A stepped structure is expected generically for \hkl(101) twin boundaries for small deviations from ideal CTB character either in misorientation or inclination. Stepped structures are not as common for symmetric tilt GBs with misorientations near the \hkl(301) CTB as shown in Fig. \ref{fig:near301}. The \hkl(807) boundary in Fig. \ref{fig:near101} is 90.2$^\circ$ in inclination space away from the \hkl(301) CTB and is an example of a well-optimized near-$90^\circ$ symmetric tilt grain boundary structure. We note that grain rotation was not allowed (misorientation was fixed) during the optimization of these GB structures.

A striking feature of the ideal $\Sigma 2$ dichromatic pattern in Fig. \ref{fig:CSL}a is the existence of short period asymmetric interfaces bounded by prismatic \hkl{101} P-type planes on one side of the interface and \hkl{100} A-type planes on the other side. In analogy to PB/BP interfaces in HCP metals, we call these facets PA/AP interfaces with the convention that the bounding plane of the twinned grain is labeled first. To simplify notation, we refer to PA/AP facets as AP facets generically unless the ordering has significance to our results. These interfaces have asymmetric \hkl[010] tilt GB character and are found to be ubiquitous features of our twin growth simulations. For the ideal $c/a$ ratio, these facets lie 150/30$^\circ$ from \hkl(301) CTBs and 120/60$^\circ$ from \hkl(101) CTBs with additional disclination character for non-ideal $c/a$ ratios. The disclination character for the MTP $c/a$ ratio is indicated by the angular deviation in the locations of A and P planes for each type of twin in Fig. \ref{fig:CSL}b-c. 

In the \hkl(301) twin growth simulations shown in Fig. \ref{fig:tgmtp}, AP facets are observed between \hkl(301) and \hkl(-101) CTB facets and are also observed between CTB facets of the same type. Short AP facets (1-2 nm) grow in length when twin boundary migration is constrained by obstacle pinning. Dumbbell-like growth structures containing AP facets form as a result of the pinning of near \hkl(-101) facets by defects in the initial metastable twin boundary inclusion in Fig. \ref{fig:tgmtp}a. AP facets are observed approximately 118-122$^\circ$ from \hkl(-101) CTB facets. In Fig. \ref{fig:tgmtp}b, AP facets form approximately 146-154$^\circ$ from \hkl(301) CTB facets. In either case, a common mechanism for the extension of AP facets appears to be the accumulation of h(2) twinning disconnections at the CTB-AP facet junction. The formation and dissociation of AP facets via the absorption and emission of twinning dislocations confers a large degree of flexibility to twin growth in $\beta$-Sn in overcoming obstacles. 

\begin{figure}[H]
\centering\leavevmode \includegraphics[width=0.9\textwidth]{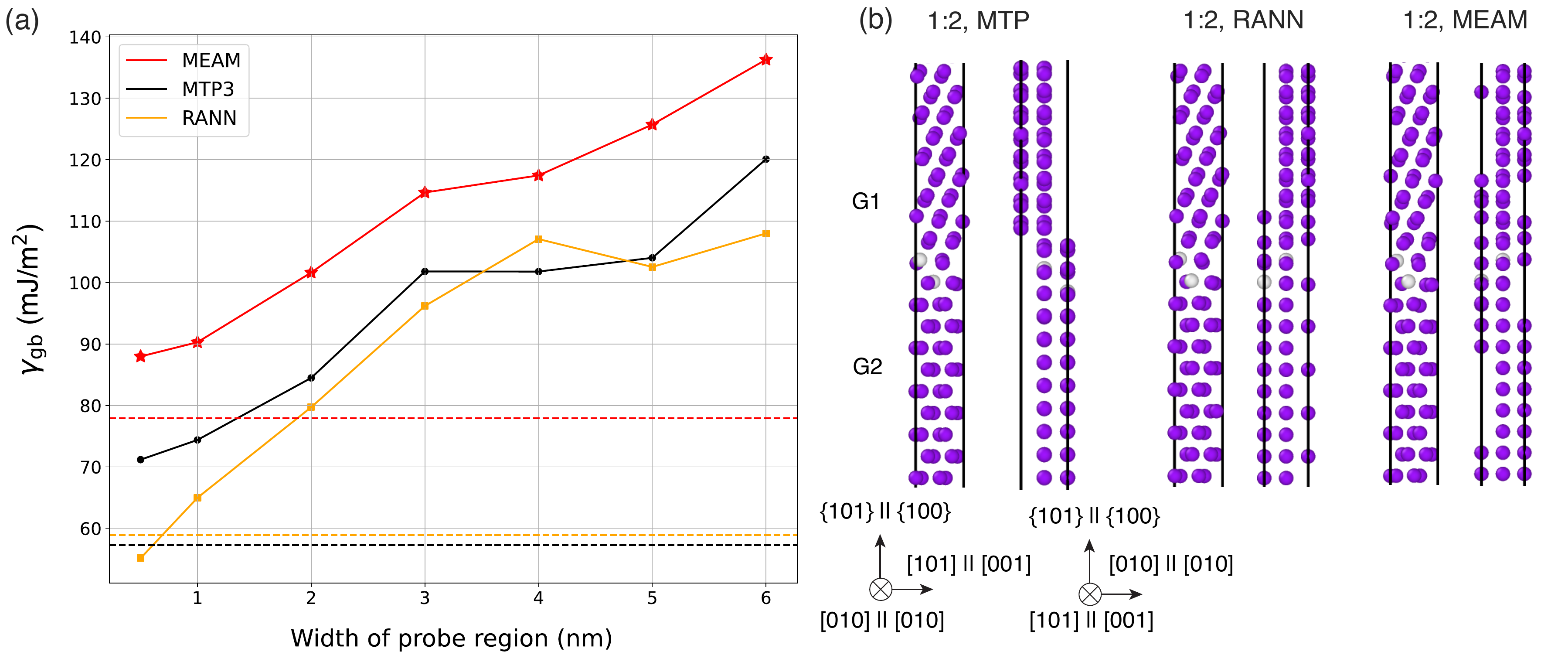}
\caption{Energy and structure of AP boundaries across potentials (a) GB energy is dependent on the size of the box used to calculate the energy because of the strain in each crystal. Dashed lines give the minimum coherent twin boundary energy for each potential. (b) Relaxed AP boundary structures are similar across potentials with two views shown. The coordinate directions follow the convention $\hkl<G1> \parallel \hkl<G2>$.}
\label{fig:AP}
\end{figure}

Coherent AP interface structures are shown in Fig. \ref{fig:AP} with energies given in Table \ref{tab:twin}. We find low AP interface energies that are consistent with the observation of these interfaces during twin growth or in the relaxed initial structures of twin inclusions. Coherent AP interfaces are created by homogeneously straining grains such that one period along the \hkl[101] direction in the P-oriented grain is matched to two periods along the \hkl[001] c-type direction in the A-oriented grain. We label this interface the 2:1 interface to distinguish between other variants which reduce misfit strain or account for disclination character. For the MTP, the misfit strain is partitioned such that a compressive strain of 3.6\% is applied along the \hkl[101] direction in the P-oriented grain and a tensile strain of 1.6\% is applied along the \hkl[001] direction in the A-oriented grain. The short period structures shown in Fig. \ref{fig:AP}b match the structure of many short AP/PA facets observed in our twin growth simulations. The relaxed structure of the 2:1 interface is not observed to vary significantly with interatomic potential. 

During twin growth simulations, misfit strain is expected to be localized at the asymmetric interface, decaying away from the interface. In our idealized interface energy calculations, elastic strains are applied homogeneously to each grain and interface energy is measured as a function of the width of a virtual probe region centered at the interface. The values reported in Table \ref{tab:twin} assume a box size of 0.5 nm which is taken as a plausible decay length for misfit strain. We note that, similar to measurements for PB interface energies in HCP metals \cite{xu2013importance}, the measured PA/AP interface energies in this work are sometimes found to be lower than those for CTBs depending on the interatomic potential. For the MTP, the AP interface energy is higher than the \hkl(301) CTB energy but slightly smaller than the \hkl(101) CTB energy. For the RANN potential, the AP interface energy is smaller than both CTB energies. For the MEAM potential, the AP interface energy is higher than that of either coherent twin. The structure of AP interfaces in twin growth simulations has additional disclination character (see Fig. \ref{fig:CSL}b-c) that is expected to raise interface energy. The study of semi-coherent AP interface structure is left to future work. 

\subsection{Comparison of twin growth behavior across potentials}
\label{subsec:twin_growth_pot}

\begin{figure}[H]
\centering\leavevmode \includegraphics[width=0.5\textwidth]{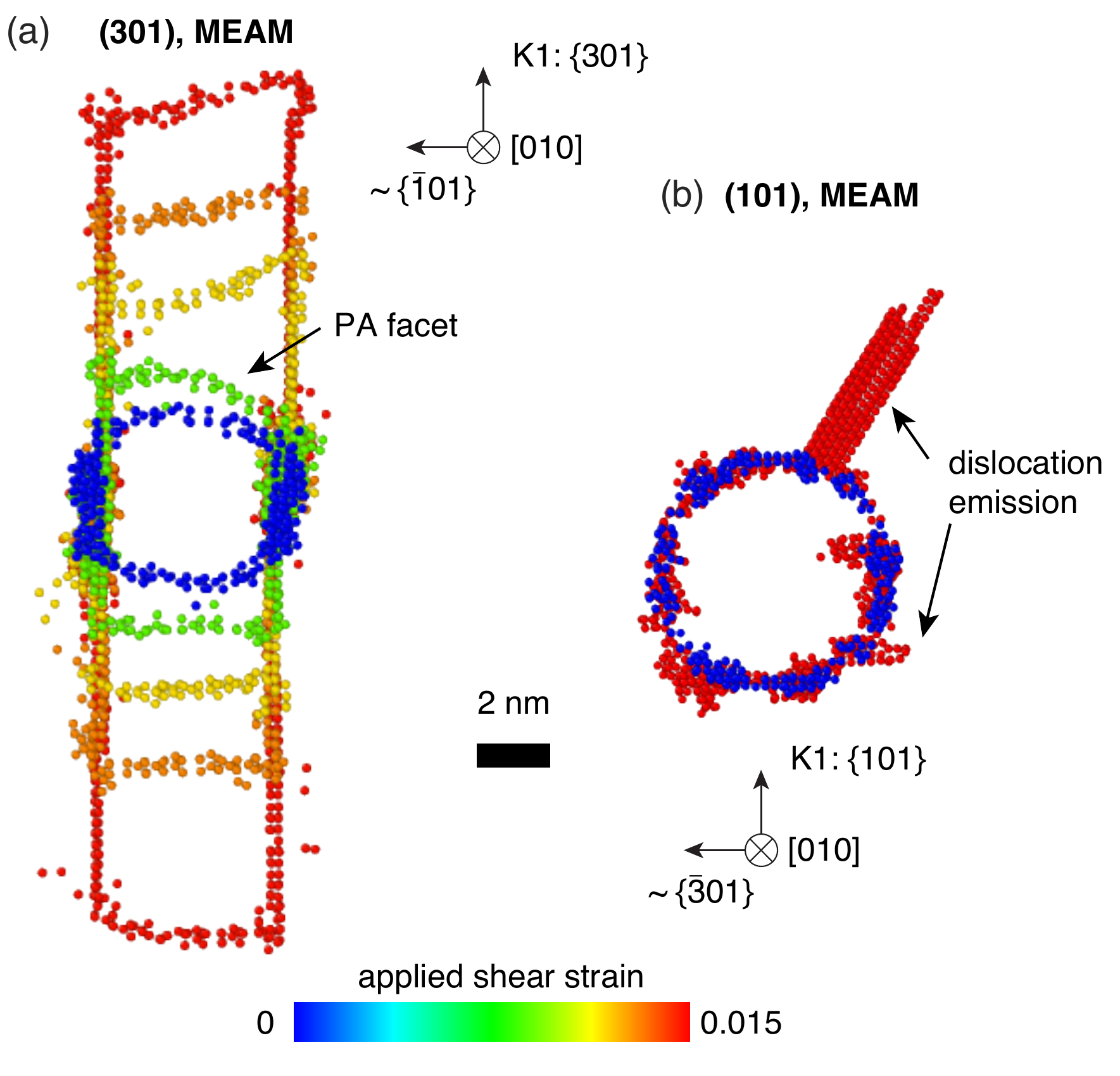}
\caption{Response of (a) \hkl(301) twin inclusion and (b) \hkl(101) twin inclusion to a shear strain resolved along CTBs using the MEAM potential at 300 K. Growth is only observed for case (a) with a vertical growth morphology that differs from the horizontal morphology observed for the MTP in Fig. \ref{fig:tgmtp}b.}
\label{fig:tgmeam}
\end{figure}

Only the ML potentials are found to exhibit twin growth over a range of experimentally plausible conditions including the growth of both predominant twin types at surveyed temperatures 10--300 K under K1 and K2 shear loading. The Achilles heel of the MEAM potential with regard to twinning is its inability to capture an h(2) disconnection mediated migration mechanism for the \hkl(101) CTB. This issue was previously discussed in the context of \hkl(101) CTB migration in Section \ref{sec:sc}, where a sliding mechanism was observed for the MEAM CTB with a large coupling factor $>1$ and a large critical stress $> 500$ MPa at 10 K. Fig. \ref{fig:tgmeam} demonstrates the limited ability of the MEAM potential to capture twin growth. An unusual vertical growth morphology is observed for \textit{K1 shear} applied to a \hkl(301) twin inclusion in Fig. \ref{fig:tgmeam}a with no motion of the near \hkl(-101) facets along the shear direction. A \textit{K1 shear} applied to a \hkl(101) twin inclusion results in dislocation emission rather than twin growth. \textit{K1 shear} loading states (not shown) also result in dislocation emission for both twin types. The MEAM potential is unable to capture the structure or migration of the experimentally observed semi-coherent near-\hkl(101) interface structures discussed in Section \ref{sec:twin_growth}. 

\begin{figure}[H]
\centering\leavevmode \includegraphics[width=0.8\textwidth]{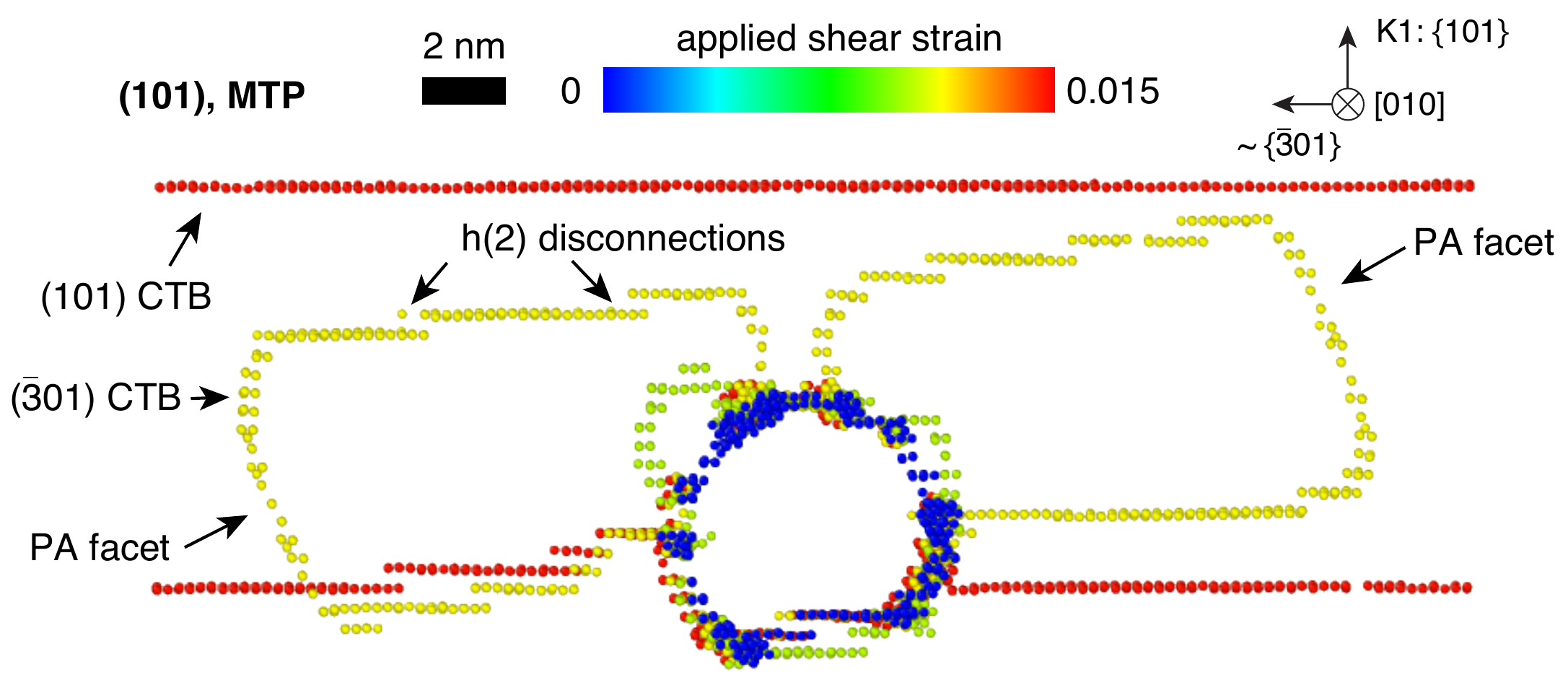}
\caption{Growth of a \hkl(101) twin inclusion using the MTP with a shear strain resolved upon the \hkl(101) CTB at 10 K.}
\label{fig:tg101}
\end{figure}

In contrast to the MEAM potential, the MTP is able to capture the growth of \hkl(101) twin inclusions. An example is shown for \textit{K1 shear} in Fig. \ref{fig:tg101}. An initially metastable twin inclusion grows via the protrusion of wing-like twin segments (shown in yellow) with both PA facets and \hkl(-301) CTB facets observed at the twin tips. After annihilation of the twin tips in the shear direction, a flat \hkl(101) CTB migrates upward while the lower \hkl(101) CTB is pinned by defects from the initial metastable twin structure. This simulation is a proof of concept for twin growth using the MTP and future work could consider more systematic structure optimization. 

\section{Discussion}
\label{sec:discussion}

Atomistic simulations have provided significant insights into twinning in cubic and HCP metals over the past few decades. To our knowledge, this work is the first direct atomistic study of twin boundary migration in a tetragonal material, $\beta$-Sn. Strong analogies are found between twinning in $\beta$-Sn and twinning in HCP metals including the observation of PA/AP facets in $\beta$-Sn similar to PB/BP facets in HCP metals  \cite{dang2020formation, xu2013importance} and the observation of double-layer h(2) twinning disconnections with small shear coupling factors \cite{gong2017interface}. These features can be rationalized by the geometry of the dichromatic pattern (Fig. \ref{fig:CSL}). However, bicrystallography alone is insufficient for quantitative predictions of twin growth kinetics and accurate interatomic potentials are needed for discovery-scale simulations of twin growth which complement high resolution experiments. 

The ML-based interatomic potentials developed in this work give results more consistent with experiments than the classical MEAM potential, including small coupling factors long thought to be associated with twinning in $\beta$-Sn \cite{tu1970direct} and observed growth of both twin types under appropriate loading conditions. A semi-coherent twin boundary morphology observed in our experiments is found to only be plausibly captured by the ML-based potentials. The MTP in particular is employed for the majority of twin growth simulations in this work because of its computational efficiency. 

\begin{figure}[H]
\centering\leavevmode \includegraphics[width=0.75\textwidth]{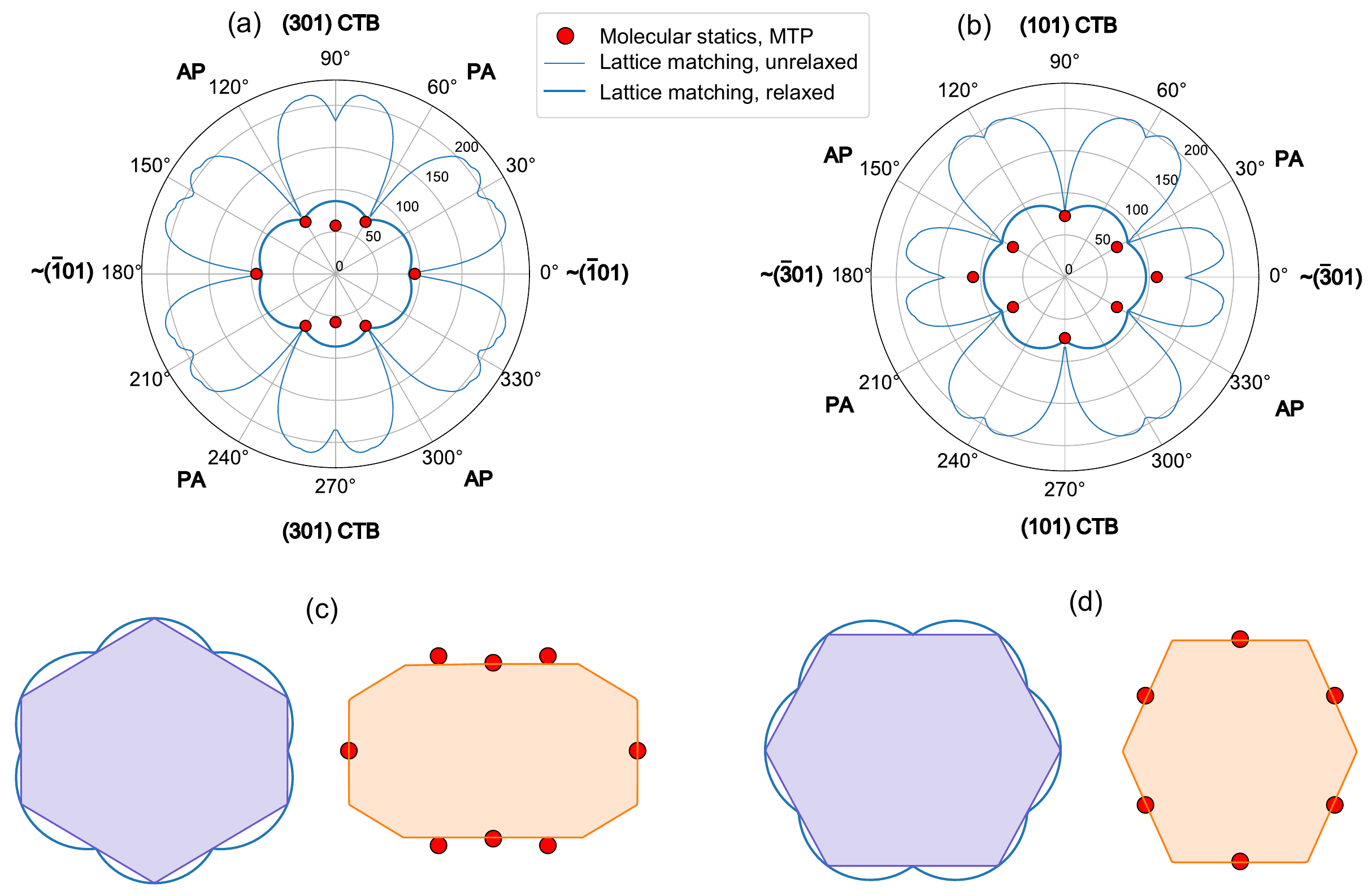}
\caption{Comparison of energies between lattice matching and MS calculations with the MTP for (a) \hkl(301) CTB and (b) \hkl(101) CTB along with predicted Wulff shapes from lattice matching and MS.}
\label{fig:wulff}
\end{figure}

This work also reveals features of twin boundary structure that are expected to be universal across all potentials surveyed in this work and observable in experiments. One such feature is the presence of local minima in the GB energy landscape at particular symmetric tilt GBs (CTBs) and asymmetric tilt GBs such as AP/PA facets as shown in Fig. \ref{fig:wulff}. The Lattice Matching (LM) method \cite{LM1,LM2} was used to compute the energy of twin-related boundaries with \hkl[010] tilt axes and misorientations corresponding to \hkl(301) and \hkl(101) twin inclusions. The interatomic potential used by the LM method is a geometry based cost function which is simpler than the classical or ML-based interatomic potentials used in this work. Cusps in the unrelaxed LM data are observed at CTBs, AP/PA facet inclinations and at 90$^\circ$ boundaries, consistent with the observation of these boundaries in our simulations of deformation twinning. MTP calculations of twin boundary energy are shown in Fig. \ref{fig:wulff} for facet inclinations observed in our simulations. 


The local minima in the GB energy landscape imply Wulff shapes expected at thermal equilibrium for quasi-2D twin inclusions in $\beta$-Sn. Wulff shapes are shown schematically for the LM and MTP data in Fig. \ref{fig:wulff}. All Wulff shapes are predicted to contain AP/PA facets. The LM method overestimates the energy of the \hkl(301) CTB compared to the MTP. Unrelaxed LM is known to systematically overestimate energy, as no atomistic relaxation is considered. A better energy estimate is obtained by applying facet relaxation to the unrelaxed LM prediction \cite{LM2}.
While this sometimes corresponds to visible faceting (e.g. the breakup of the \hkl(301) CTB into AP/PA facets), it is alternatively interpreted as a reduced-order relaxation mechanism that approximates the effect of atomistic relaxation. Our atomistic calculations and prior experiments \cite{kaira2016microscale} imply the existence of atomically sharp non-faceted  \hkl(301) CTBs in the structure of \hkl(301) twin inclusions. Regardless of differences in energies between LM and MS, the existence of deep cusps in the GB energy landscape across potentials at AP/PA facet inclinations suggests a high likelihood of observing these boundaries in $\beta$-Sn microstructures. 

\begin{figure}[H]
\centering\leavevmode \includegraphics[width=0.7\textwidth]{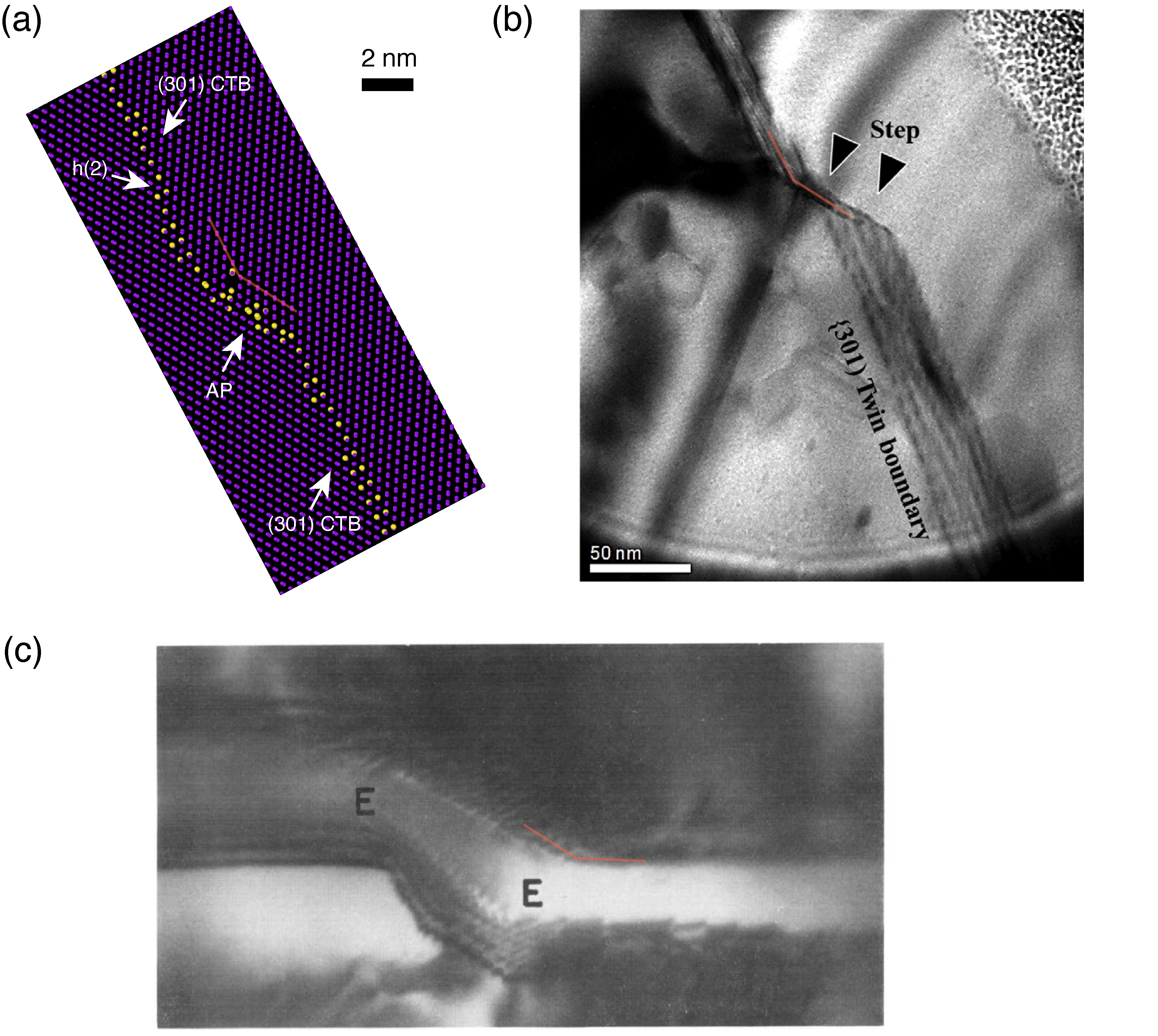}
\caption{Comparison of experimental micrographs of \hkl(301) twins containing likely AP facets to simulation data. (a) shows a typical AP facet observed in this work using the MTP (b) shows a facet resulting from twin boundary-dislocation interactions \cite{kaira2016microscale} and (c) a solidification twin  \cite{fourie1960growth} with \hkl(301) CTBs separated by a region described as incoherent between E and E. The 150$^\circ$ angular marker in red is superimposed upon each image.}
\label{fig:exptAP}
\end{figure}

We identify likely PA/AP interfaces from prior experimental studies of \hkl(301) twins \cite{kaira2016microscale, fourie1960growth} in Fig. \ref{fig:exptAP}, which were observed but were not explicitly characterized. Fig. \ref{fig:exptAP}b shows a simulation snapshot from a MTP simulation with shear stress resolved along a \hkl(301) CTB at 200 K where an AP facet has formed along the CTB. In the simulation, the AP facet resides at an angle of approximately 150$^\circ$ with respect to the CTB, consistent with the ideal angle in the $\Sigma2$ CSL (Fig. \ref{fig:CSL}a). The angular marker in red in Fig. \ref{fig:exptAP}a fits the experimental data in Fig. \ref{fig:exptAP}b-c reasonably well. Atomic resolution in the TEM data would be needed to make more precise statements about AP facet angles as TEM contrast alone can be misleading. 

There are several notable differences among the twin boundary facet structures from simulation and experiments. The length of the experimentally observed near-AP facets (around 50 nm) are larger than those in our simulations by an order of magnitude. Though scale bars were not reported in \ref{fig:exptAP}c \cite{fourie1960growth}, a TEM magnification of 200,000X was given which we estimate corresponds to a scale bar of 50 nm similar to \ref{fig:exptAP}b. The observation of relatively long AP facets in experiments is similar to observations of $\sim$50 nm BP facets in HCP twins \cite{dang2020formation}, an observation which was initially surprising given the disclination character of BP facets. The disclination character of AP/PA or BP/PB facets is expected to lead to an accumulating Burgers vector content with increasing facet length that must be accommodated in the twin microstructure \cite{el2015, dang2020formation}. In the case of BP facets in HCP materials, observed accommodation mechanisms in MD simulations include the formation of misfit dislocation networks and emitted stacking faults \cite{dang2020formation, yue2023twin}. We note that, in our simulations, stacking faults are not observed to be required to form short segments of PA/AP interfaces. These segments form via the nucleation and coalescence of twinning disconnections, similar to short BP/PB facets in HCP twins \cite{gong2017interface}. Larger twin growth simulations could be conducted in future work to discover as of yet unknown accommodation mechanisms for the formation of large AP/PA facets.

\section{Conclusion}
\label{sec:conclusion}

We have studied the structure and migration of twin boundaries in tetragonal $\beta$-Sn via a combination of experiments, atomistic simulations and bicrystallography-based methods. Two ML-based potentials were developed in this work (using the MTP and RANN formalisms) which enabled simulations of twin growth in $\beta$-Sn in agreement with experiments. The following conclusions can be drawn from this work: 

\begin{enumerate}
\item Three types of low energy interfaces in $\beta$-Sn are found to be ubiquitous during twin growth: the symmetric \hkl(301) and \hkl(101) CTBs and asymmetric coherent AP/PA facets with \hkl{100} (A) and \hkl{101} (P) planes joined at the interface. All three interfaces are characterized by deep cusps in the GB energy landscape and are expected facets in the Wulff shapes of \hkl(301) or \hkl(101)-type twins. 

\item AP/PA facets were observed, but not characterized, in prior experiments. AP/PA facets form in our twin growth simulations as a result of both disconnection pile-ups and energy minimization, similar to BP/PB facets in HCP metals. 

\item CTB motion for both twin types is found to be mediated by the nucleation and propagation of h(2) disconnections with steps spanning two interatomic planes. Identical negative coupling factors of around -0.1 were found for the \hkl(301) and \hkl(101) CTBs using ML-based potentials, consistent with predictions from experiments and crystallography-based models. The MEAM potential predicts a positive coupling factor $>1$ for the \hkl(101) CTB, restricting its ability to accurately capture twinning. All potentials predict a larger critical stress for migration of the \hkl(101) vs. \hkl(301) CTB. 

\item An unexpected growth morphology was observed for a \hkl(301) twin in our experiments with a long axis defined by a semi-coherent near-\hkl(101) interface rather than the expected \hkl(301) CTB. The semi-coherent interface has $\sim$ 30 nm \hkl(101) CTB segments separated by $\sim$ 2-3 nm steps with dislocation character. Such a morphology can be explained by the relaxation of groups of twinning disconnections and by a loading state that favors growth of  the near-\hkl(101) twin boundary facet rather than the \hkl(301) CTB. Similar morphologies were observed in our direct MD simulations with a lower degree of relaxation compared to experiments. 

\end{enumerate}

This work provides fundamental data such as interface orientations, energies, coupling factors and critical stresses that can be used to inform mesoscale models for twinning as well as multiphase strength models for $\beta$-Sn. The ML-based potentials developed in this work are expected to be applicable to diverse microstructural evolution problems in Sn involving extended defects, twinning and phase transformations. 

Our simulations primarily focused on deformation twinning at low temperatures in idealized quasi-2D systems. The combination of bicrystallography-based and atomistic methods used in this work lays the groundwork for fully 3D studies of twin boundary structure and growth in $\beta$-Sn. We hope that the present work encourages further experimental characterization of twin related defects and facet structures in $\beta$-Sn. The character of AP/PA type interfaces has yet to be directly confirmed with atomic resolution. We are not aware of any experimental micrographs showing AP facets along \hkl(101) CTBs, in which case a shallow CTB-AP facet angle around 120$^\circ$ is predicted. $90^\circ$ facet structures in \hkl(101) CTBs also await experimental characterization. Little is still known about twin nucleation and detwinning mechanisms during high strain rate deformation \cite{PTW2025} or the impact of alloying on twin boundary structure and migration.

\section*{Acknowledgments}

IC acknowledges support from the Los Alamos National Laboratory (LANL) Postdoc Program via the Metropolis Fellowship and acknowledges computational resources from the LANL Advanced Simulation and Computing Program.
DNB and KD acknowledge support from the Institute for Material Science's Rapid Response program at Los Alamos National Laboratory in the early stages of this work.
 AH acknowledges support from the Physics and Engineering Models (PEM) Materials project within the Advanced Simulation and Computing (ASC) Program at LANL. We would like to thank Doyl Dickel for a fruitful discussion regarding interatomic potential development and Calvin Lear for experimental guidance. 

\section*{Data Availability}
All interatomic potentials (MTP and RANN), training databases, and sample LAMMPS 
input structures and scripts used in this study are available here: \\\url{https://gitlab.com/mtp_potentials/sn_ml_potentials}. 

\FloatBarrier
\bibliographystyle{utphys-custom}

\bibliography{citations}

\newpage

\appendix
\renewcommand{\thesection}{S}
\renewcommand{\thefigure}{S\arabic{figure}}
\renewcommand{\thetable}{S\arabic{table}}
\setcounter{figure}{0}
\setcounter{table}{0}

\section{Supplementary Information}

\subsection{Interatomic potentials}

\paragraph{Training database}

A diverse DFT training database was generated to fit the ML potentials. Multiple allotropes of Sn ($\alpha$, $\beta$, 
BCT, body-centered cubic (BCC)-Sn) and two non-observed 
phases (simple cubic (SC) and FCC) were considered along with the liquid phase. Structural perturbations were considered including hydrostatic and uniaxial strains (up to 10\% in magnitude), thermal perturbations, point defects and free surfaces. Supercells of various sizes were randomly 
distorted by atomic displacements of 0.1--0.5~\AA  \ corresponding to an effective temperature range of 100--1200~K. Additional details of database generation 
can be found in the authors' previous studies \cite{nitol2022machine,Nitol:2023}. 

\paragraph{Selected properties}
Table~\ref{tab:beta-sn} compares experimental data, DFT calculations, and predictions 
obtained from MTP and RANN for several key structural and elastic parameters of 
\texorpdfstring{$\beta$}{beta}-Sn. The equilibrium lattice parameters 
($a$, $c$), volume, and elastic constants ($C_{11}$, $C_{12}$, $C_{13}$, $C_{33}$, 
$C_{44}$) are all listed.

\begin{table*}[ht]
    \centering
    \caption{Key material properties of \texorpdfstring{$\beta$}{beta}-Sn: equilibrium 
    lattice constants ($a$, $c$ in \AA), volume (in \AA$^3$/atom), and selected elastic 
    constants ($C_{ij}$ in GPa). Experimental data and DFT results are also provided 
    for comparison.}
    \label{tab:beta-sn}
    \resizebox{\textwidth}{!}{%
    \begin{tabular*}{\textwidth}{@{\extracolsep{\fill}} lcccccccc} 
    \hline
         & $a$ & $c$ & Vol/atom  & $C_{11}$ & $C_{12}$ &  $C_{13}$ & $C_{33}$ & $C_{44}$ \\
    \hline
    Expt.\cite{gale2003smithells} & 5.83 & 3.18 & 28.37 & 73.2 & 59.8 & 39.1 & 90.6 & 21.9 \\
    DFT  & 5.95  & 3.21  & 28.37 & 69   & 45   & 30   & 82   & 14   \\
    MTP  & 5.97  & 3.22  & 28.74 & 64   & 44   & 27   & 80   & 6    \\
    RANN & 5.95  & 3.21  & 28.36 & 59   & 61   & 27   & 89   & 8    \\
    MEAM \cite{Ko:2018} & 5.86  & 3.21 & 27.51  & 90 & 47  & 37  & 94  & 8  \\
    \hline
    \end{tabular*}
    }
\end{table*}

Although the two ML potentials capture the main features of the 
\texorpdfstring{$\beta$}{beta}-Sn structure, some deviations from DFT are observed in certain elastic constants, particularly $C_{44}$ which is softer than experimentally measured values across all potentials. 

\begin{figure}[!htbp]
    \centering
    \includegraphics[width=0.8\textwidth]{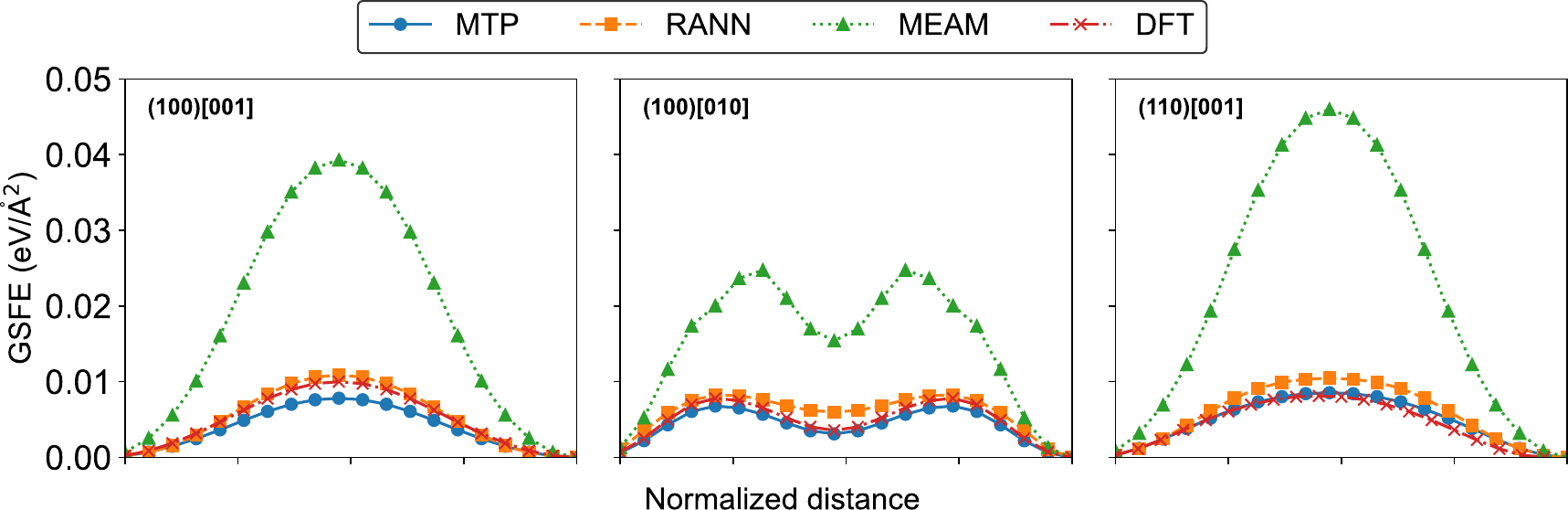}
    \caption{Stacking fault energies for $\beta$-Sn for different slip systems. The DFT results are from \cite{Nitol:2023}. These DFT results are not included in the training set.} 
    \label{fig:GSFE}
\end{figure}

Only considering the basic properties in Table \ref{tab:beta-sn}, it is not clear which potential should capture twinning most accurately in $\beta$-Sn relative to experiments. A known advantage of the previously fit ML potential for Sn \cite{Nitol:2023} (the predecessor of the ML potentials in this work) is the more accurate stacking fault energies for common slip systems in $\beta$-Sn compared to MEAM. Fig. \ref{fig:GSFE} shows that both the MTP and RANN potentials developed in this work capture stacking fault energies in closer agreement to DFT data than the MEAM potential for three types of slip systems. 

\paragraph{MTP parameter optimization:}
In the MTP framework, the cutoff radius ($r_c$) and the maximum level of expansion 
($\text{lev}_{\text{max}}$) are critical parameters that govern the balance between 
accuracy and efficiency. A grid search was performed in this work, with 
$\text{lev}_{\text{max}}$ scanned from 18 to 22 and $r_c$ scanned from 5.0~\AA\ to 
6.0~\AA\ (in increments of 0.1~\AA). For each pair of parameter values, 10 training 
runs were initiated with random initial conditions, producing a total of 300 MTP 
models. The best compromise between accuracy and computational cost was obtained with 
$\text{lev}_{\text{max}} = 18$ and $r_c = 5.0$~\AA.

\paragraph{RANN network architecture:}
In the RANN formalism, a single hidden layer containing 23 neurons was used to model 
the interatomic interactions in Sn. The resulting network architecture for $\beta$-Sn 
is $29 \times 12 \times 1$, where 29 descriptors are transformed into 12 hidden 
neurons, and a single output neuron provides an atomic energy contribution. Several 
meta-parameters were set as follows:$m \in \{0,\dots,5\}, n \in \{-1,\dots,3\}, r_e = 2.880350,\alpha = 4.897871,\beta_k \in \{1,2,4,8\},C_{\min} = 0.4,C_{\max} = 1.6$

\begin{figure}[!htbp]
    \centering
    \includegraphics[width=0.5\textwidth]{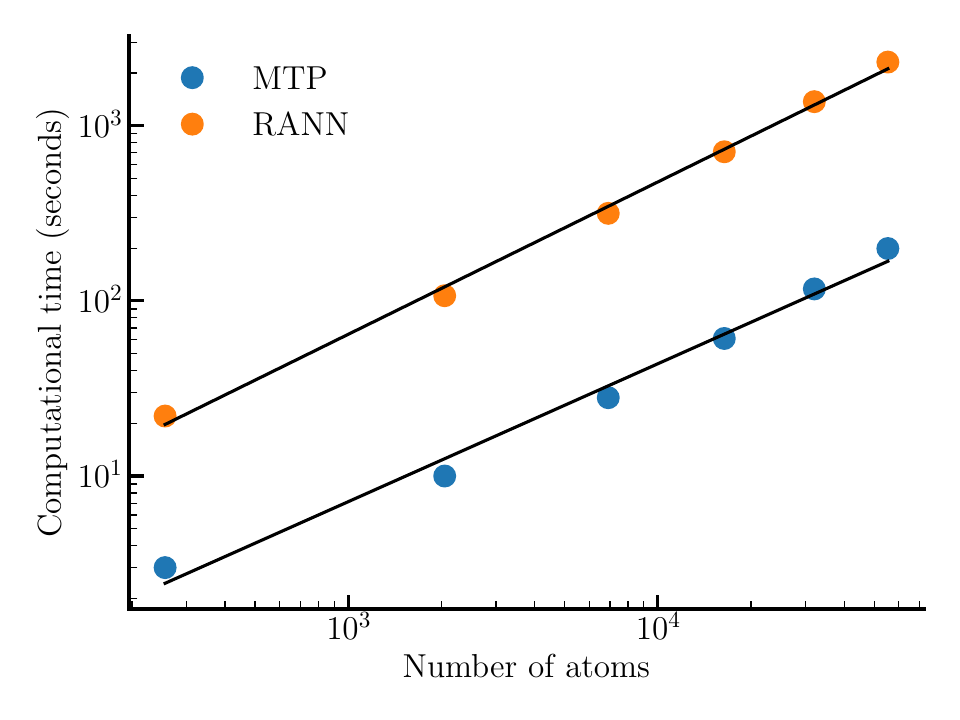}
    \caption{Computational time versus the number of atoms using MTP  and RANN potentials.} 
    \label{fig:comp_time}
\end{figure}

\paragraph{Accuracy and computational speed}
For RANN, the root-mean-square error (RMSE) on the training set was 1.88 meV/atom, and the validation set was higher at 2.05~meV/atom. For MTP, the RMSE was 
6.62 meV/atom on energies and 0.05 eV/\AA\ on forces. Fig. \ref{fig:comp_time} shows the total computational time (in seconds) versus the number of atoms for single-point calculations using both  MTP (blue circles) and RANN (orange circles) models. The data were obtained on a single compute node with 128 processors. Although both methods exhibit roughly linear scaling with respect to the total number of atoms, MTP consistently demonstrates lower overall computational times compared to RANN. The vertical spacing between the two curves represents the per-system speed differences directly, highlighting that MTP offers better performance across the tested range of system sizes. For the twin migration simulations conducted in this work, the MTP simulations were only around two times slower than the MEAM simulations. 

\begin{figure}[H]
\centering\leavevmode \includegraphics[width=0.7\textwidth]{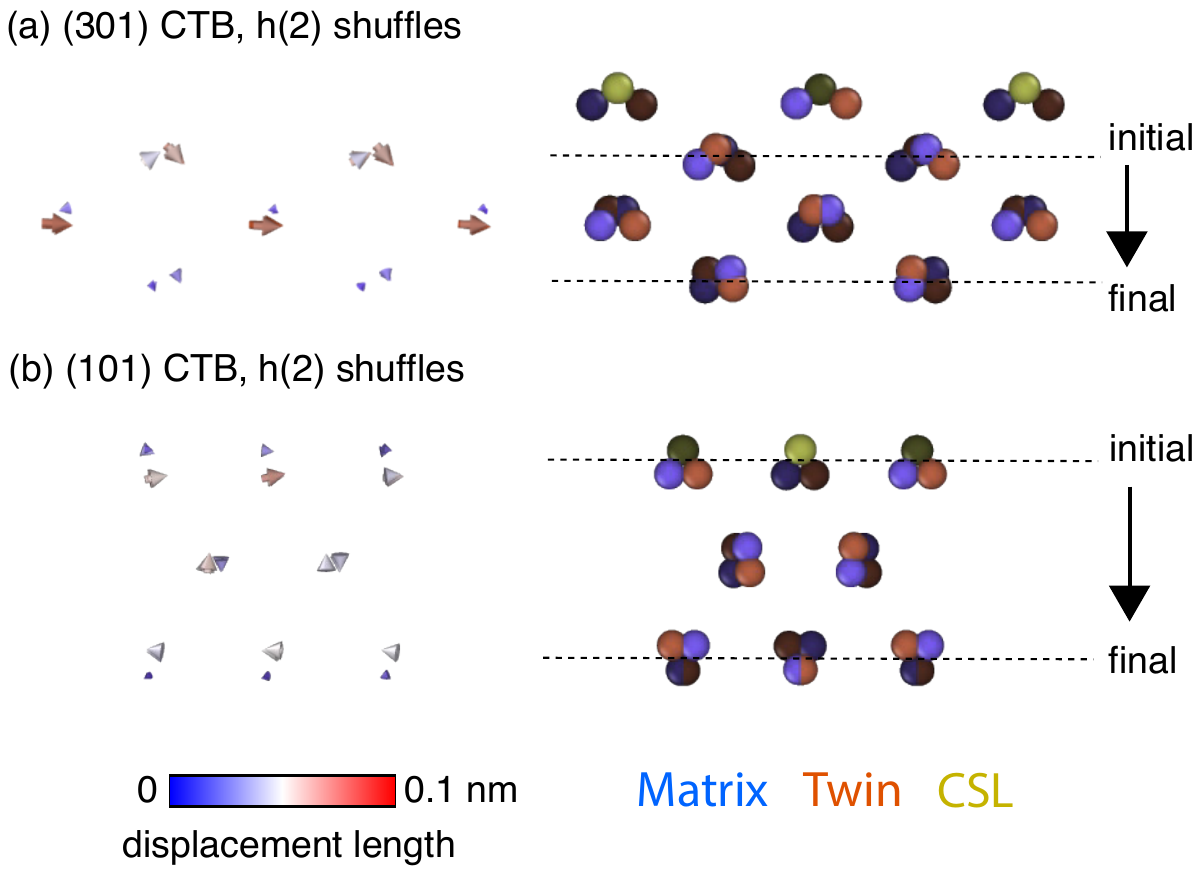}
\caption{Illustration of shuffles swept out during passage of an h(2) disconnection for (a) (301) CTB and (b) (101) CTB with atomic displacements from MD MTP trajectories at 10 K shown on the left and coherent dichromatic patterns shown on the right. Displacements are colored by length. Shuffles are found to be distance minimizing in the dichromatic pattern.}
\label{fig:shuffle}
\end{figure}

\subsection{Additional bicrystallography of twin boundaries in \texorpdfstring{$\beta$-Sn}{beta-Sn}}

h(2) disconnection motion is accompanied by three layers of atomic shuffles as shown in the MD MTP trajectories for \hkl(301) and \hkl(101) CTBs in Fig. \ref{fig:shuffle}. These shuffling patterns are examples of the X mechanism discussed by Christian \textit{et al.} \cite{christian2002} in the general context of shuffles for 2-lattice structures. Layers of shuffle displacements in the vertical direction alternate between having primarily shear character and normal character. Shuffles are distance minimizing in the dichromatic pattern. This observation is consistent with prior analysis of low barrier shear coupled migration mechanisms in FCC metals \cite{chesser2021optimal}.

The oiLAB software package \cite{oiLAB} was used to explore the bicrystallography of twin boundaries in $\beta$-Sn. Conflicting reports of $\Sigma$ values exist in the literature for \hkl(301) and \hkl(101) type twins \cite{kaira2016microscale, gupta2022electrodeposition}. We show in Table \ref{tab:sigma_theta_values} that the value of $\Sigma$ depends on the rational approximant used for the c/a ratio. A 3D CSL only exists for rational values of $(c/a)^2$. The $\Sigma$ values for $c/a = \sqrt{8/27}$ (close to the experimentally reported value) match those of a study of GB character distribution in electrodeposited Sn \cite{gupta2022electrodeposition}. The $\Sigma$ value has possible significance to twin growth in $\beta$-Sn because it sets the width of strain-free inclusions. Stated differently, an incoherent twin facet 90$^\circ$ from the CTB must span $\Sigma$ planes in order to achieve a strain free structure. Fourie \textit{et al.} \cite{fourie1960growth} reported that \hkl(301) twin growth occurred in increments of 34 \hkl(301) planes. 

\begin{table}[h!]
\centering
\begin{tabular}{|c|c|c|c|c|}
\hline
$c/a$ & $\Sigma_{301}$ & $\Sigma_{101}$ & $\theta_{301}$ (°) & $\theta_{101}$ (°) \\
\hline
$\sqrt{1/3} = 0.577$   & 2  & 2  & 60      & 120      \\
$\sqrt{3/10} = 0.548$  & 37 & 13 & 62.648  & 122.579  \\
$\sqrt{8/27} = 0.544$  & 33 & 35 & 62.964  & 122.878  \\
$\sqrt{2/7} = 0.535$   & 25 & 9  & 63.896  & 123.749  \\
\hline
\end{tabular}
\caption{Values of $\Sigma$ and misorientation angle $\theta$ for each twin type for different rational approximants of $(c/a)^2$}
\label{tab:sigma_theta_values}
\end{table}

To gain additional insight into faceted twin structures, we enumerate possible facet inclinations within the CSL for the case $c/a = \sqrt{2/7} = 0.535$ which closely approximates the MTP $c/a = 0.538$. Facet inclination angles and CSL periodicities along the facet are given for \hkl(301) and \hkl(101) twin inclusions in Tables \ref{tab:301twin} and \ref{tab:101twin}. Lengths are normalized such that $a = 1$ and can be uniformly rescaled by the actual $a$ value of interest. Only short facets with normalized length less than 30 are listed. For both twin types, CTB facets are observed as the shortest period facets with the period of the \hkl(101) CTB less than the \hkl(301) CTB, consistent with our atomistic simulations. Multiple near-AP facets are enumerated around $30^\circ$ from the \hkl(301) CTB or $60^\circ$ from the \hkl(101) CTB which have normal vectors $n_A$, $n_B$ close to A:\hkl(100) and P:\hkl(101) planes. These variants illustrate that there are multiple ways to accommodate the deviation from ideal AP character in a periodic manner. Atomistic simulations would need to be performed to determine the energies of near-AP facet variants.  

In addition to CTBs, $90^\circ$ symmetric tilt GB facets and near-AP facets, we see that short period facets exist around $60^\circ$ from the \hkl(301) CTB or $30^\circ$ from the \hkl(101) CTB with inclinations close to (301) and C:\hkl(001) planes. We call these C3 facets. C3 facets comprise shallow cusps in the unrelaxed LM energy data reported in the main text (Fig. 14), but were not directly observed in our deformation twinning simulations. Similar to near-AP facets, multiple short period near-C3 facet variants exist in the CSL. 

\begin{table}[h!]
\centering
\begin{tabular}{|l|c|c|c|c|}
\hline
Boundary Type & Inclination (°) & Period ($a=1$) & $n_A$ & $n_B$ \\
\hline
CTB              & 0       & 1.8898   & \hkl(3 0 -1)   & \hkl(3 0 1)   \\
AP               & 28.126  & 15.0000  & \hkl(28 0 -1)  & \hkl(14 0 13) \\
AP               & 29.926  & 28.347   & \hkl(53 0 -1)  & \hkl(1 0 1)   \\
AP               & 31.948  & 13.363   & \hkl(1 0 0)    & \hkl(11 0 12) \\
C3               & 58.052  & 25.000   & \hkl(42 0 11)  & \hkl(0 0 1)   \\
C3               & 61.874  & 8.017    & \hkl(13 0 4)   & \hkl(-1 0 16) \\
90° STGB         & 90.000  & 7.071    & \hkl(7 0 6)    & \hkl(-7 0 6)  \\
\hline
\end{tabular}
\caption{Facet inclinations and period lengths for $\Sigma 25$ (301) twin with $c/a = \sqrt{2/7}$ and $a = 1$.}
\label{tab:301twin}
\end{table}

\begin{table}[h!]
\centering
\begin{tabular}{|l|c|c|c|c|}
\hline
Boundary Type & Inclination (°) & Period ($a=1$) & $n_A$ & $n_B$ \\
\hline
CTB              & 0       & 1.134    & \hkl(1 0 -1)   & \hkl(1 0 1)    \\
C3               & 28.126  & 9.000    & \hkl(14 0 -5)  & \hkl(0 0 1)    \\
C3               & 29.303  & 26.005   & \hkl(41 0 -14) & \hkl(1 0 -26)  \\
C3               & 29.926  & 17.008   & \hkl(3 0 -1)   & \hkl(1 0 -17)  \\
AP               & 118.126 & 4.8107   & \hkl(5 0 4)    & \hkl(1 0 0)    \\
AP               & 121.020 & 19.803   & \hkl(19 0 17)  & \hkl(37 0 1)   \\
AP               & 120.454 & 24.608   & \hkl(8 0 7)    & \hkl(46 0 1)   \\
90° STGB         & 90.000  & 4.243    & \hkl(7 0 2)    & \hkl(7 0 -2)   \\
near-90° STGB               & 87.800  & 29.720   & \hkl(50 0 13)  & \hkl(16 0 -5)  \\
\hline
\end{tabular}
\caption{Facet inclinations and period lengths for $\Sigma 9$ (101) twin with $c/a = \sqrt{2/7}$ and $a = 1$.}
\label{tab:101twin}
\end{table}

Finally, oiLAB allows for the enumeration of disconnection modes with glide and climb character for each twin-related GB as was done for asymmetric tilt GBs in FCC metals in \cite{joshi2024}. In this work, the glide disconnection modes output for the CTBs by oiLAB are found to be consistent with the h(2) disconnections observed in simulations employing ML-based potentials.

\subsection{Addition atomistic data for twin boundary structure and energy}


\begin{figure}[H]
\centering\leavevmode \includegraphics[width=0.9\textwidth]{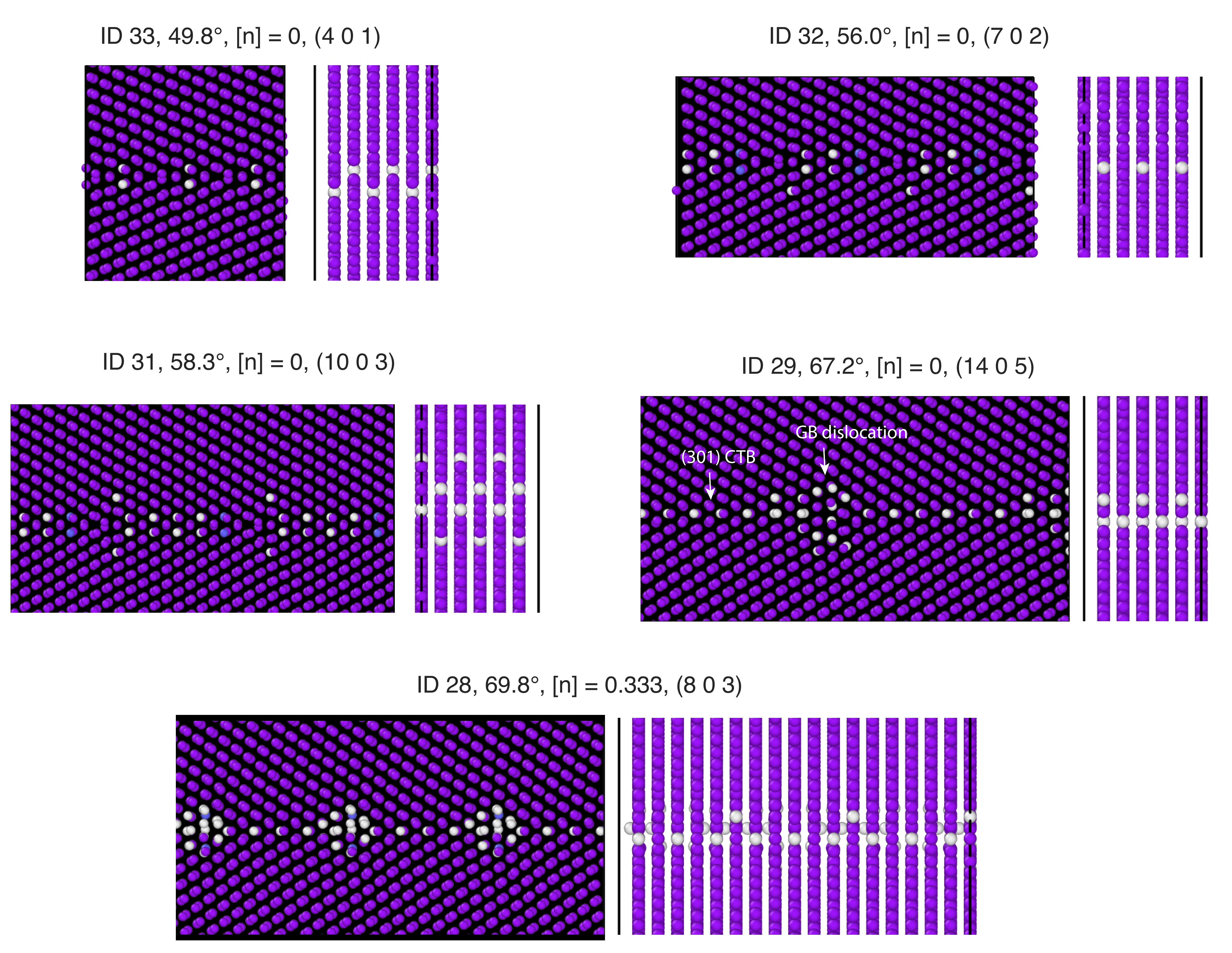}
\caption{Minimum energy symmetric tilt boundary structures with misorientations close to the \hkl(301) twin, $\theta = 63.57^\circ$. The \hkl(14 0 5) boundary, ID 29, is 95.3$^\circ$ away from the \hkl(-101) CTB in inclination space.}
\label{fig:near301}
\end{figure}

\begin{figure}[H]
\centering\leavevmode \includegraphics[width=0.9\textwidth]{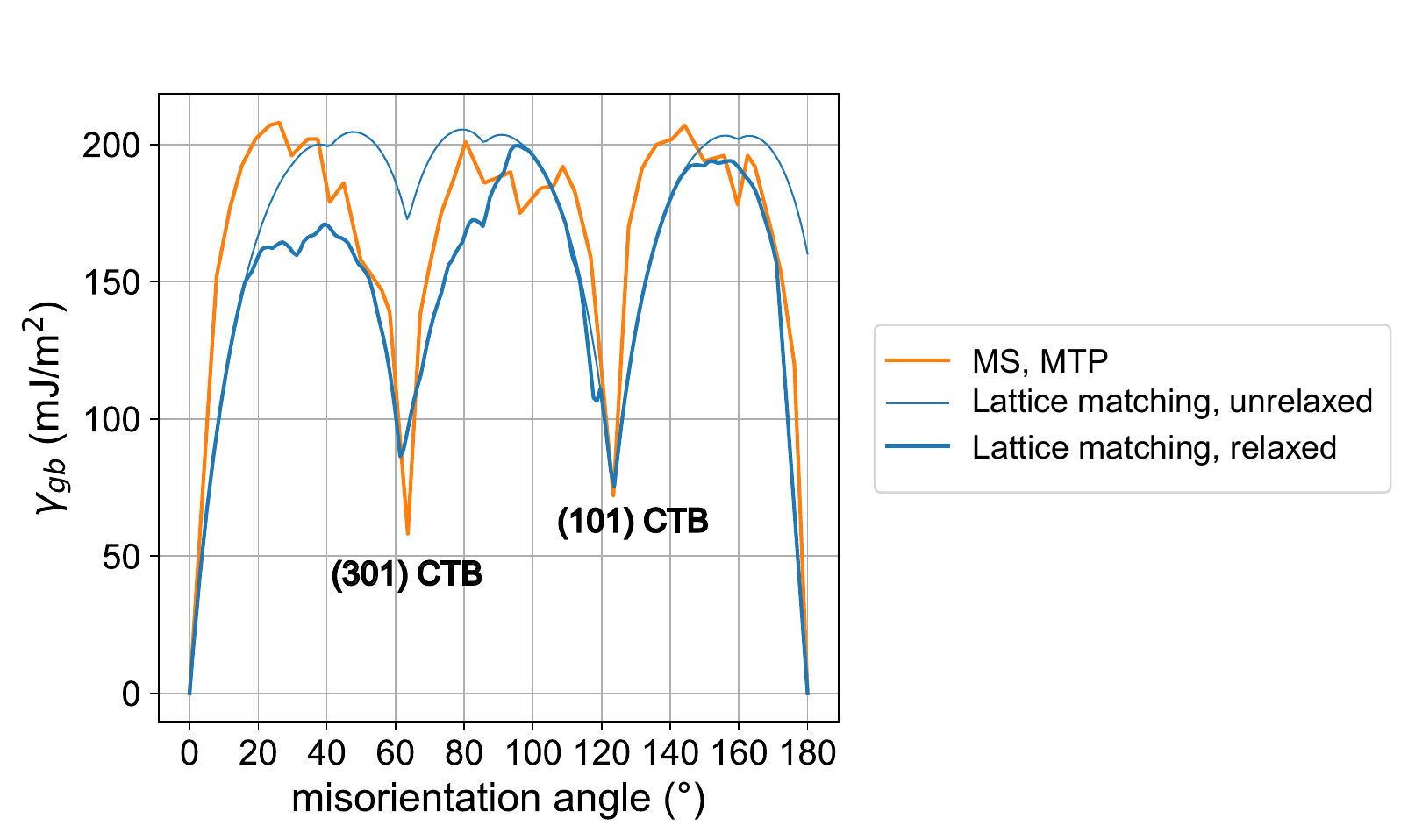}
\caption{Comparison of lattice matching energies and MS energies (MTP) for \hkl[010] symmetric tilt grain boundaries. Local energy minima at misorientations corresponding to the two types of CTBs are captured by both methods.}
\label{fig:LM_STGB}
\end{figure}


\end{document}